\documentclass[twocolumn]{aa}
\usepackage{graphicx}
\usepackage{epstopdf}
\usepackage[varg]{txfonts}
\usepackage{natbib}
\usepackage{url}
\usepackage{multirow}
\begin{document} 

\title{Multiwavelength analysis of three\\
   SNe associated with GRBs observed by GROND}

   \author{F.\,Olivares\,E.\inst{1}\fnmsep\inst{2}\fnmsep\thanks{\email{f.olivares.e@gmail.com}}
     \and J.\,Greiner\inst{1} \and P.\,Schady\inst{1} \and
     S.\,Klose\inst{3} \and
     T.\,Kr{\"u}hler\inst{4}\fnmsep\thanks{Present address: European
       Southern Observatory, Alonso de C\'{o}rdova 3107, Vitacura,
       Casilla 19001, Santiago 19, Chile.} \and A.\,Rau\inst{1} \and
     S.\,Savaglio\inst{1,5} \and D.\,A.~Kann\inst{1,3} \and
     G.\,Pignata\inst{2} \and J.\,Elliott\inst{1}\!\and
     A.\,Rossi\inst{3}\fnmsep\thanks{Present address: INAF-IASF
       Bologna, Area della Ricerca CNR, via Gobetti 101, 40129
       Bologna, Italy.}\!\and M.\,Nardini\inst{6}\!\and
     P.\,M.\,J.\,Afonso\inst{7}\!\and R.\,Filgas\inst{8}\!\and
     A.\,Nicuesa\,Guelbenzu\inst{3}\!\and S.\,Schmidl\inst{3}\!\and
     V.\,Sudilovsky\inst{1}}

   \institute{Max-Planck-Institut f\"ur extraterrestrische Physik,
     Giessenbachstra\ss{}e 1, 85740 Garching, Germany \and
   %2:
     Departamento de Ciencias Fisicas, Universidad Andres Bello,
     Avda.\ Republica 252, Santiago, Chile \and
   %3:
     Th\"uringer Landessternwarte Tautenburg, Sternwarte 5, 07778
     Tautenburg, Germany \and
   %4:
     Dark Cosmology Centre, Niels Bohr Institute, University of
     Copenhagen, Juliane Maries Vej 30, 2100 Copenhagen, Denmark \and
   %5:
     Physics Department, University of Calabria, Arcavacata, 87036
     Rende, Italy \and
   %6:
     Universit\`a degli studi di Milano-Bicocca, Piazza della
     Scienza 3, 20126, Milano, Italy \and
   %7:
     American River College, Physics and Astronomy Dpt., 4700 College
     Oak Drive, Sacramento, CA 95841, USA \and
   %8:
     Institute\,of\,Experimental\,and\,Applied\,Physics,
     Czech\,Technical\,University\,in\,Prague,
     Horska\,3a/22,\,12800\,Prague\,2,\,Czech\,Republic}

\date{Received May 20, 2013; accepted January 30, 2015}

  \abstract
  {After the discovery of the first connection between GRBs and SNe
    almost two decades ago, tens of SN-like rebrightenings have been
    discovered and about seven solid associations have been
    spectroscopically confirmed to date.}
  {The luminosity and evolution origin of three SN rebrightenings in
    GRB afterglow light curves at $z\sim0.5$ are determined along with
    accurate determinations of the host-galaxy extinction. Physical
    parameters of the SN explosions are estimated, such as synthesised
    $^{56}$Ni mass, ejecta mass, and kinetic energy.}
  {We employ GROND optical/NIR data and \emph{Swift} X-ray/UV data to
    estimate the host-galaxy extinction by modelling the afterglow
    SED, to determine the SN luminosity and evolution, and to
    construct quasi-bolometric light curves. The latter were corrected
    for the contribution of the NIR bands using data available in
    the literature and blackbody fits. Arnett's analytic approach has
    been employed to obtain the physical parameters of the explosion.}
  {The SNe 2008hw, 2009nz, and 2010ma observed by GROND exhibit 0.80,
    1.15, and 1.78 times the optical ($r'$ band) luminosity of SN
    1998bw, respectively. While SN 2009nz exhibits an evolution
    similar to SN 1998bw, SNe 2008hw and 2010ma show earlier peak
    times. The quasi-bolometric light curves (340\,--\,2200 nm)
    confirm the large luminosity of SN 2010ma ($1.4\times10^{43}$ erg
    s$^{-1}$), while SNe 2008hw and 2009nz reached a peak luminosity
    closer to SN 1998bw. The modelling resulted in $^{56}$Ni masses of
    around $0.4\text{\,--\,}0.5\,\text{M}_\sun$.}
  {By means of the a very comprehensive data set, we found that the
    luminosity and the $^{56}$Ni mass of SNe 2008hw, 2009nz, and
    2010ma resembles those of other known GRB-associated SNe. This
    findings strengthens previous claims of GRB-SNe being brighter
    than type-Ic SNe unaccompanied by GRBs.}


   \titlerunning{Multiwavelength analysis of three GRB-SNe}
   \authorrunning{Olivares~E.~et al.}

   \maketitle

%________________________________________________________________

\graphicspath{{./figs/}}

\section{Introduction}\label{c1_Intro}

Gamma-ray bursts (GRBs) and supernovae (SNe) correspond to the most
energetic explosions in the Universe with a radiative energy release
of about $10^{\,51\mbox{--}53}$\,erg. Nowadays the observational
evidence points towards the catastrophic deaths of massive stars,
which are thought to give birth to both long GRBs \citep[durations
  $\ga 2$ s;][]{Kouveliotou93} and broad-lined (BL) type-Ic SNe after
the collapse of their cores into a black hole
\citep[BH;][]{Paczynski98a,FryWooHar99,vanParadijs+00}. Known as the
collapsar model \citep{Woosley93,MacWoo99,Bromberg+12}, the collapsing
core of a very massive star can lead to the formation of a
relativistic jet that will produce high-energy emission
\citep{Woosley93,WooMac99} in the form of a GRB or an X-ray flash
\citep[XRF;][]{Heise01,Kippen04,Sakamoto+08}. The $\gamma$-ray
emission itself lasts from a few tenths of a second to a few thousand
seconds, is generated within the outflow at ultra-relativistic
velocities, and is collimated into a jet \citep{Zhang09} that drills
its way out of the star. The interactions between fireball shells with
different speeds (``internal shocks'') are responsible for the prompt
$\gamma$-ray emission. The multi-wavelength afterglow (AG), detectable
from radio throughout to X-rays up to months after the GRB
\citep[e.g.,][]{Kann10,Kann11}, is explained by the synchrotron
emission produced in the interaction between the circumburst material
and the relativistic outflow \citep[``external shocks''; see][for a
  review]{ZhangMeszaros04}.

In principle, the energy transferred to the envelope should also be
capable of causing the ejection of the stellar envelope
\citep{Burrows00,Heger+03}. However, it is unclear how or even if
there is always enough energy for the SN explosion
\citep[``fall-back'' events, e.g.,][and references
  therein]{Fryer+07}. Moreover, it is unknown exactly to what extent
the progenitors have to lose their envelope to produce a GRB. However,
it is generally accepted that type-Ib and type-Ic SNe are formed from
evolved high-mass progenitors like Wolf-Rayet (WR) stars, which have
liberated their outer shells through (1) pre-SN stellar winds, (2)
mass transfer to a binary companion due to Roche-lobe overflow, or (3)
a combination of both processes.  The stellar explosion is then
referred to as a ``stripped-envelope'' SN \citep[SE
  SN;][]{ClocchiattiWheeler97}. The end result of a GRB-SN explosion
would correspond to a compact remnant, either a neutron star
\citep[NS;][]{BaadeZwicky34} or a BH \citep{Arnett96}. To date, long
GRBs have been associated only with type-Ic BL SNe, which are those
lacking H \citep{Minkowski41} and He lines \citep{Filippenko97} and
showing expansion velocities in the order of 20,000 km s$^{-1}$ (for
reviews on the GRB-SN connection, see \citealt{WoosleyBloom06} and
\citealt{HjorthBloom11}).

%% OBSERVATIONAL BACKGROUND & SPECTROSCOPY %%
The first and most representative case of the GRB-SN connection is the
association of \object{SN 1998bw} with the under-luminous \object{GRB
  980425} \citep{Kippen98a,Sadler98}. Although initially controversial
\citep{Galama98b,Pian+98}, the physical association between these
events was supported on temporal \citep{Iwamoto+98} and spatial
grounds \citep{Pian+00,Kouveliotou04}. Five years later, the
association of \object{GRB 030329} with \object{SN 2003dh} was clearly
identified through spectra showing both the AG and SN counterparts
\citep{Hjorth03,Kawabata03,Stanek03,Matheson03} and became a solid
piece of evidence in favour of the GRB-SN connection. There have been
a number of other spectroscopic associations\footnote{These also
  showed BL~features~in~their~\mbox{spectra}:~GRB 021211/SN
  2002lt\,\citep{DellaValle03},\,GRB\,031203/SN\,2003lw\,(Malesani
  et~al.~2004),~GRB 050525A/SN 2005nc~\mbox{\citep{DellaValle06a},
    GRB} 060218/SN\,2006aj\,(\citealt{Pian+06,Modjaz+06};\,Sollerman
  \mbox{et~al.~2006),~GRB\,081007/SN\,2008hw\,(\citealt{DellaValle08};
    Jin et~al.}
  2013),~GRB~091127/SN~2009nz~\mbox{\citep{Berger+11}},~GRB
  101219B/SN\,2010ma\,\citep{Sparre+11a},\,GRB\,111211A\,(de\,Ugarte
  Postigo
  et~al.~2012),\,GRB\,100316D/SN\,2010bh\,\citep{Chornock10c,Bufano+12},\,GRB
  120422A/SN 2012bz\,\citep{Melandri+12,Schulze+14},\mbox{GRB
    120714B/SN 2012eb\,\citep{Klose+12a,Klose+12b}},
  \mbox{GRB\,130215A \citep{deUgarte+13a,Cano+14},\,GRB} 130427A/SN
  2013cq \citep{Xu+13}, GRB 130702A/SN 2013dx \citep{Schulze+13}, GRB
  130831A/SN 2013fu \citep{Klose+13,Cano+14}, and GRB 140606B
  \citep{Perley+14}}, which in the literature are also dubbed
``hypernovae'' \citep[HNe;][]{Paczynski98b,Hansen99} given their high
luminosities. Due to their high energetics, HNe produce
$\ga0.2\,\text{M}_\sun$ of $^{56}$Ni, are thought to have very massive
progenitors, and are often connected to BH formation
\citep{Nomoto10,Stritzinger+09}.

%% LATE-TIME BUMPS AND SAMPLES %%
Late-time rebrightenings in AG light curves have been interpreted as
SN signals, e.g., GRBs 970228 \citep{Galama00a,ReiLamCas00}, 011121
\citep{Bloom+02b,Greiner03}, 020405 \citep{Price03,Masetti03}, 041006
\citep{Stanek05,Soderberg06}, 060729, and 090618 \citep[the latter two
  in][]{Cano11b} to mention a few. These photometric bumps are
consistent in terms of colour, timing, and brightness with those
expected for the GRB-SN population
\citep{ZehKloseHartmann04,Ferrero06}, but they are usually at faint
apparent magnitudes, which hampers the spectroscopic
identification. However, the SN counterpart can be as bright as
$M_V=-19.8$ mag for SN 2003lw \citep{Malesani04}. These rebrightenings
have been detected in AG light curves out to redshifts of $\sim1$
\citep[e.g.,][]{Masetti05,DellaValle03} owing to the sensitivity of
current ground-based telescopes dedicated to follow-up observations. A
handful of sample studies of GRB-SNe (including bumps not
spectroscopically identified) have analysed the luminosity
distribution, the light-curve morphology, and the explosion physical
parameters such as kinetic energy ($E_\text{k}$), ejected mass
($M_\text{ej}$), and $^{56}$Ni mass
\citep[$M_\text{Ni}$;][]{Richardson09,Thoene11,Cano13}. They concluded
that GRB-SNe are in general brighter than the local sample of SE SNe,
except for cases of exceptionally bright type-Ic SNe (e.g., SN 2010ay,
\citealt{Sanders11}; SN 2010gx, \citealt{Pastorello+10}).  Regarding
light-curve morphology, \citet{Stanek05} and more recently
\citet{Schulze+14} claim to have found a correlation between
brightness and light-curve shape, which was also confirmed by
\citet{Cano14} using a larger sample and including the appropriate K
corrections. This strengthens the use of GRB-SNe as standard candles
for cosmology (see also recent studies by \citealt{LiHjorth14},
\citealt{CanoJakobsson14}, and \citealt{Li+14}). While more than two
dozen photometric bumps in AG light curves have been presented as SN
rebrightenings \citep[e.g.,][]{Richardson09}, so far only seven have
been confirmed through high-S/N spectra: SNe 1998bw, 2003dh, 2003lw,
2006aj, 2010bh, 2012bz, 2013dx, and 2013cq.

%% NON-DETECTIONS %%
% While no SN signature is expected from short GRBs, thought to be
% produced by mergers of compact objects (for deep non-detections, see
% the cases of GRBs 050509B in \citealt{Hjorth05a} and
% \citealt{Bloom+06}, and 050709 in \citealt{Fox05} and
% \citealt{Hjorth05b}), there are a few supposedly long events where
% an expected SN appearance was never detected. In the cases of GRBs
% 060505 \citep{Fynbo06,Ofek07} and 060614
% \citep{Fynbo06,GalYam06,DellaValle06b}, there are very tight
% constraints on the SN signature, which go down to 1\% as bright as
% SN 1998bw. The validity of the SN non-detections \citep[see][for a
% comprehensive summary]{Kann11} as a classification tool is still a
% point of controversy \citep[e.g.,][]{Zhang09}.

%% MAGNETARS %%
The energy injection of a newly-formed rapidly-spinning
strongly-magnetised NS (so-called ``magnetar'') provides an
alternative scenario for GRB-SNe. Here the SN is powered by the
dipole-field strength of the magnetar
\citep[e.g.,][]{Woosley10,Dessart+12}. Magnetars have been linked to
the GRB emission too, because their outflows can explain the
energetics of long-duration GRBs
\citep[e.g.,][]{Bucciantini+09,Metzger+11}.  Moreover, the
$E_\text{k}$ of GRB-SNe ($\sim 10^{52}$ erg) is fairly consistent with
the maximum rotational energy of a NS with a period of 1 ms
\citep{Mazzali+06b}. GRB-SN zoo is claimed to be entirely produced by
magnetars and driven by the SN rather than by the GRB jet
\citep{Mazzali+14}.

%% BRIDGE %%
Three detected SNe associated with GRB counterparts are the main focus
of this paper: SNe 2008hw (GRB 081007), 2009nz (GRB 091127), and
2010ma (GRB 101219B). The acquisition, reduction, and calibration of
the multiwavelength data are described in Sect.\ \ref{c2_Data}. The
corresponding analysis is presented in Sect.\ \ref{c3_4Reb} along with
further discussion in Sect.\ \ref{c4_Diss}. Finally, we summarise our
conclusions in Sect.\ \ref{Summary}.

\section{Data}\label{c2_Data}

For the three objects of interest, the~data~was~obtained~by~the
X-Ray\,Telescope\,\citep[XRT;][]{Burrows05} and the Ultra-Violet
Optical Telescope\,\citep[UVOT;][]{Roming05}~both~on board the
\emph{Swift} satellite \citep{Gehrels04} and by the Gamma-Ray burst
Optical and Near-infrared Detector
\citep[GROND;][]{Greiner07,Greiner08}, the seven-channel imager
mounted on the MPG 2.2-m telescope at La Silla, Chile. The whole data
set comprises X-ray photometry and spectra from $0.2\mbox{\,--\,}10$
keV, UV/optical photometry in the $uvw2\,uvm2\,uvw1\,u\,b\,v$ filters,
and optical/NIR photometry in the
$g\,'\,r\,'\,i\,'\,z\,'\,J\,H\,K_\text{s}$ bands, spanning four orders
of magnitude in the energy spectrum.

The UVOT/XRT data retrieval and the GROND/UVOT methodology towards the
final photometry are detailed in \citet{Olivares+12}. Optical image
subtraction of the host galaxy was performed for GRB 081007/SN 2008hw
and GRB 091127/SN 2009nz. All data presented are corrected for the
Galactic foreground reddening $E(B-V)_\text{Gal}$ in the direction of
the burst \citep{Schlegel+98}. The reddening is transformed to the
extinction $A_{V,\text{Gal}}$ by assuming a ratio of total to
selective absorption of $R_{V,\text{Gal}}=3.1$ from the Milky-Way (MW)
reddening law. The final GROND photometry is presented in Appendix
\ref{app1}. All magnitudes throughout the paper are in the AB system.

\section{Three GRB-associated SNe detected by GROND}\label{c3_4Reb}

\begin{table*}
  \caption{GROND sample of GRB-associated SNe.}
  \label{tSample}
  \centering
% \begin{tabular*}{1.1\textwidth}{@{\extracolsep{\fill}}llcccccccl}
  \begin{tabular*}{\textwidth}{@{\extracolsep{\fill}}llcccccccl}
    \hline\hline\noalign{\vspace{0.5\smallskipamount}}
    GRB  &SN &RA(J2000) &Dec.(J2000) &$z$ &$d$\,\tablefootmark{a} &$A_{V,\text{Gal}}$\tablefootmark{b} &$N_\text{H,Gal}$\tablefootmark{c} &Refs. \\
    &   &$[\ ^\mathrm{h}\ $:$\ ^\mathrm{m}\ $:$\ ^\mathrm{s}$ ] &[ $^\circ$ : $'$ : $''$ ]  &  &[Mpc]  &[mag]  &[$10^{20}$\,cm$^{-2}$] & \vspace{0.5\smallskipamount} \\
    \hline\noalign{\vspace{0.5\smallskipamount}}
    081007  &2008hw  &22:39:50.40 &$-40$:08:48.8 &$0.530$   &2885  &$0.05$  &$1.4$   &1  \\
    091127  &2009nz  &02:26:19.87 &$-18$:57:08.6 &$0.490$   &2628  &$0.12$  &$2.8$   &2  \\
    101219B &2010ma  &00:48:55.35 &$-34$:33:59.3 &$0.552$   &3022  &$0.06$  &$3.1$   &3  \\
    \hline
  \end{tabular*}
  \tablefoot{\tablefoottext{a}{The luminosity distances are computed
      using the $\Lambda$CDM cosmological model
      \citep[$\Omega_\text{M}=0.27$, $\Omega_\Lambda=0.73$, and
        $H_0=74.2$ km s$^{-1}$ Mpc$^{-1}$;][]{Riess09} and the
      redshifts corrected by the local velocity field
      \citep{Mould00}.}  \tablefoottext{b}{The Galactic foreground
      extinction values are taken from the dust maps of
      \citet{Schlegel+98}.}  \tablefoottext{c}{The absorption column
      densities are taken from the Galactic \ion{H}{i} maps of
      \citet{Kalberla05}.} } \tablebib{Redshifts are taken from (1)
    \citealt{Berger+08}, (2) \citealt{Vergani+11}, (3)
    \citealt{Sparre+11b}.  }
\end{table*}
%%label:tSample

\begin{table*}
  \caption{GROND photometry of the host galaxies.}
  \label{tHosts}
  \centering
% \begin{tabular*}{1.2\textwidth}{@{\extracolsep{\fill}}llccccccc}
  \begin{tabular*}{\textwidth}{@{\extracolsep{\fill}}llccccccc}
    \hline\hline\noalign{\vspace{0.5\smallskipamount}}
    GRB  &SN &$g\,'$ &$r\,'$ &$i\,'$ &$z\,'$ &$J$ &$H$ &$K_\text{s}$ \vspace{0.5\smallskipamount} \\
    \hline\noalign{\vspace{0.5\smallskipamount}}
    081007  &2008hw &$24.66\pm0.11$ &$24.49\pm0.11$ &$24.08\pm0.19$ &$23.96\pm0.24$ &$>22.0$ &$>21.1$ &$>20.1$ \\
%    &$\cdots$ &$\cdots$ &$\cdots$ &$\cdots$ &$22.07\pm0.26$ &$21.19\pm0.33$ &$21.79\pm0.78$ \vspace{1.5\smallskipamount} \\
    091127  &2009nz &$24.08\pm0.09$ &$23.45\pm0.06$ &$22.85\pm0.07$ &$22.95\pm0.08$ &$>21.7$ &$>21.4$ &$>19.9$ \\
%    &$\cdots$ &$\cdots$ &$\cdots$ &$\cdots$ &$21.88\pm0.90$ &$22.12\pm0.60$ &$\cdots$ \vspace{1.5\smallskipamount} \\
    101219B &2010ma &$>25.4$ &$>25.2$ &$>24.5$ &$>24.5$ &$>22.2$ &$>22.0$ &$>20.2$ \\
%    &$\cdots$ &$25.35\pm0.25$ &$\cdots$ &$\cdots$ &$23.4\pm2.0$ &$22.5\pm1.4$ &$20.7\pm1.1$ \\
    \hline
  \end{tabular*}
  \tablefoot{The host-galaxy magnitudes are all corrected for the corresponding Galactic foreground
    extinction. The upper limits were derived from the deepest
    observation available showing no detection and are quoted at the
    3$\sigma$ confidence level.}
\end{table*}
%%label:tHosts

Table \ref{tSample} presents a sub-sample of GRBs with late-time
optical SN rebrightenings in their AG light curves, all of them
observed by GROND. Deep late-time observations were carried out for
each of them to constrain the contribution from their host
galaxies. If the host was detected, we performed image
subtraction. Table \ref{tHosts} presents the resulting photometry for
those host galaxies. In the following, observational facts and general
properties of each event are summarised from the literature. If
possible, mass estimates are derived from the SEDs of the host
galaxies using the \texttt{hyperZ} code \citep{Bolzonella+00} and a
library of galaxy spectral templates extinguished by the different
reddening laws.

\vspace{-1mm}\paragraph{\bf GRB 081007/SN 2008hw} The \emph{Swift}/BAT
\citep{Barthelmy05} discovered GRB 081007 at 05:23:52 UT on 2008
October 7 \citep{Baumgartner+08}. The prompt emission had a duration
of $T_{90}\approx10$ s and a soft spectrum with $E_\text{peak}\la30$
keV \citep{Markwardt+08}. The redshift of $z=0.5295$ was found by
\citet{Berger+08} through optical spectroscopy. A subsequent optical
spectrum taken 17 days after the burst shows broad features indicative
of an emerging SN, which was thereafter classified as type I (no
Hydrogen lines) and named SN 2008hw \citep{DellaValle08}. The SN bump
was also reported as a flux excess with respect to the AG
\citep{Soderberg+08b}.  The GROND photometry of the host galaxy (Table
\ref{tHosts}) from August 31, September 30, and October 21, 2011,
yields a stellar-mass range of
$M_\star\sim10^{8\mbox{\,--\,}9}\,\text{M}_\sun$, which is compatible
with the population of GRB hosts \citep{Savaglio+09}. Using
appropriate transformation
equations\footnote{\url{http://www.sdss.org/dr7/algorithms/sdssUBVRITransform.html}}
our host magnitudes are somewhat brighter but marginally consistent
with the measurements published by \citet{Jin+13} of $R_C>24.67$ and
$I_C=24.29\pm0.20$ mag at $\sim87$ d after the GRB.

\vspace{-1mm}\paragraph{\bf GRB 091127/SN 2009nz} At 23:25:45 UT on
2009 November 27, the \emph{Swift}/BAT was triggered by GRB 091127
\citep{Troja+09}. The $\gamma$-ray emission lasted for $T_{90}=7.1$ s
and showed a soft spectrum \citep{Stamatikos+09,Troja+12}. A redshift
of $z=0.490$ was obtained from optical spectroscopy
\citep{Cucchiara+09,Thoene+09}. Observations by Konus-Wind confirmed
the results from the \emph{Swift}/BAT \citep{Golenetskii+09} and
additionally yielded an energy release typical for cosmological GRBs
($E_{\gamma,\text{iso}}\sim10^{52}$ erg). The optical AG was confirmed
with GROND observations \citep{Updike+09} adding NIR detections. The
full analysis of the GROND AG light curve was presented in
\citet{Filgas+11b}. The SN classification became official based on the
photometric SN bump \citep{Cobb+10c,Cobb+10a} and spectroscopy was
published later showing BL features \citep{Berger+11}. Photometry
depicting the SN rebrightening was published in \citet{Cobb+10b} and
\citet{Vergani+11}. Using the host-galaxy detections in GROND optical
imaging (October 31, 2010) and in NIR photometry from
\citet{Vergani+11}, a stellar mass of
$M_\star=10^{8.4}\,\text{M}_\sun$ is obtained. This value falls in the
low-mass end of the observed distribution of GRB host masses
\citep{Savaglio+09} and is compatible with the stellar mass computed
by \citet{Vergani+11}.

\vspace{-3mm}\paragraph{\bf GRB 101219B/SN 2010ma} At 16:27:53 UT on
2010 December 19, the \emph{Swift}/BAT discovered GRB 101219B
\citep{Gelbord+10}. The BAT burst lasted $T_{90}\simeq34$ s
\citep{Cummings+10} and consisted of a spectrum with
$E_\text{peak}\simeq70$ keV as observed by \emph{Fermi}/GBM
\citep{vanderHorst10}. The SN discovery was first reported
photometrically by \citet{Olivares11} along with a redshift estimation
assuming the brightness of SN 1998bw for the rebrightening
($z=0.4\mbox{\,--\,}0.7$). The spectroscopic confirmation of SN 2010ma
came later by \citet{Sparre+11a} along with the redshift determination
of $z=0.55185$ from weak Mg absorption lines. The spectroscopy lead to
further analysis by \citet{Sparre+11b} that shows broad-line features
characteristic of GRB-SNe. Late-time GROND observations on September
30, 2011, show no signal of a host-galaxy down to deep limits (Table
\ref{tHosts}), therefore no image-subtraction procedure was
performed. These upper limits imply a stellar mass for the host galaxy
of $M_\star\la10^{9.2}\,\text{M}_\sun$, which corresponds to the
low-mass half of observed GRB host mass distribution and is marginally
compatible with the Small Magellanic Cloud (SMC).

\subsection{Multicolour light-curve fitting}\label{s4_LCs}
After image subtraction of the host galaxy in the cases where it was
detected (Table \ref{tHosts}), the light curves were fitted
simultaneously using one or two power-law components ($F_\nu\sim
t^{\,\alpha}$) and templates of SN 1998bw, where corrections due to
redshift and Galactic foreground extinction were taken into account
\citep[see][for details on the fitting of SN 1998bw
  templates]{ZehKloseHartmann04}. Simultaneous modelling consists of
unique power-law slopes $\alpha_1$, $\alpha_2$, and SN-template
stretch factor $s$ for all bands. The ratio between the luminosity of
the observed SN and that of SN 1998bw \citep[luminosity ratio
  $k$;][]{ZehKloseHartmann04} represents the free brightness
parameter, which was fitted to the light curves corrected for Galactic
extinction only. Therefore, the luminosity ratios are then corrected
for the host-galaxy extinction $A_{V,\text{host}}$ determined by the
SED modelling (see Sect.\ \ref{s4_SEDs}). The modelling is described
in detail below for each event and summarised in Tables \ref{tAGpars}
and \ref{tSNpars}.

\begin{figure}
  \centering
  \includegraphics[bb=7 26 592 739,width=\linewidth,clip]{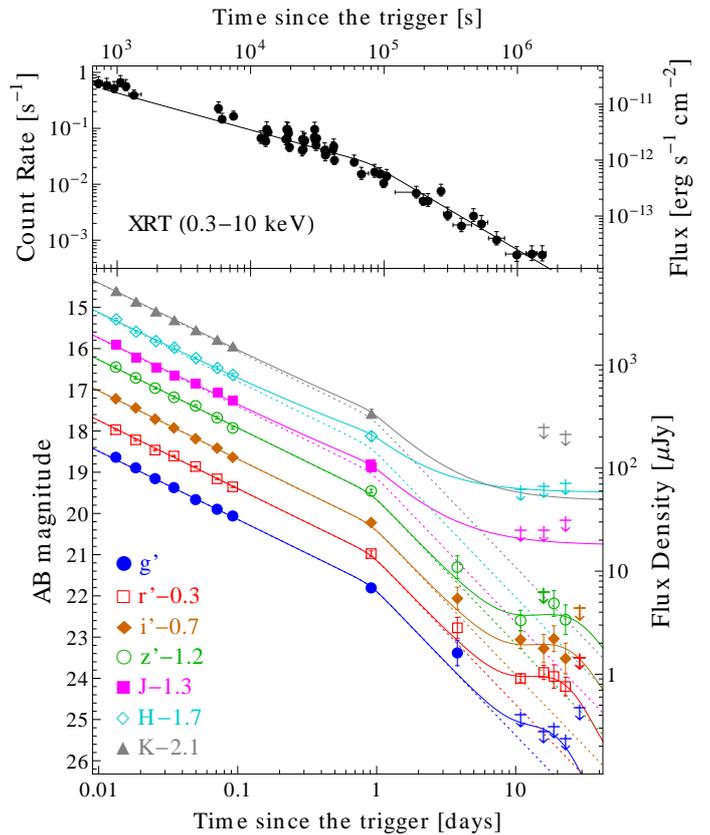}
  \caption{Multicolour light curves of GRB 081007/SN 2008hw corrected
    for Galactic extinction as observed by the \emph{Swift}/XRT (upper
    panel) and GROND (lower panel). Filled circles represent
    detections and arrows are upper limits. Solid lines correspond to
    the overall fits and dotted lines to the AG component. For clarity,
    light curves were shifted along the magnitude axis. Shallow upper
    limits are not shown (see Table \ref{t081007} for the complete
    data set).}
  \label{fLC08hw}
\end{figure}

\vspace{0mm}\paragraph{\bf GRB 081007/SN 2008hw} Figure \ref{fLC08hw}
shows that the light curves in all seven bands are well modelled using
a broken power law \citep{Beuermann+99}. The $g\,'r\,'i\,'z\,'\!$
photometry has been image-subtracted to remove the host-galaxy
flux. The X-ray light curve from the \emph{Swift}/XRT was included in
the fitting to constrain the decay after the break, where there is
only a single optical epoch. For the $r\,'i\,'z\,'\!$ bands, it was
necessary to add a supernova component with a luminosity about
$65\text{\,--\,}80$\% that of SN 1998bw (see Table \ref{tSNpars}). Due
to the $JHK_\text{s}$ flux excesses with respect to the broken power
law at roughly 1 day after the burst, a constant component was
included in the modelling for these bands. The $g\,'\!$-band upper
limit is strongly affected by absorption of metal lines and wavelength
extrapolation of the SN 1998bw template (e.g., the case of SN 2009nz
due to high redshift). \citet{Jin+13} report a luminosity 50\% that of
SN 1998bw, however, without accounting for the significant host
extinction (see Sect.\ \ref{s4_SEDs}).

\begin{figure}
  \centering
  \includegraphics[bb=4 3 597 584,width=\linewidth,clip]{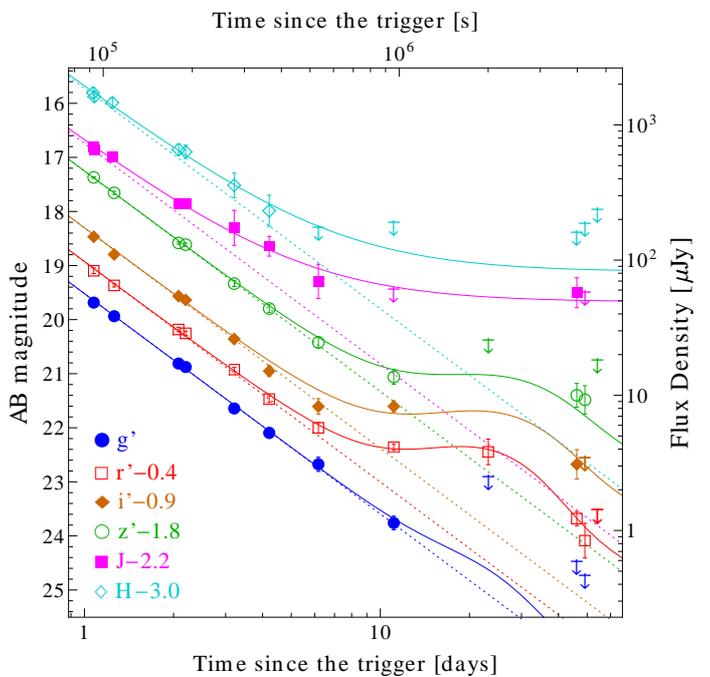}
  \caption{Multicolour GROND light curves of GRB 091127/SN 2009nz
    corrected for Galactic extinction. Only data after day one has
    been employed. The symbol and line coding is the same as
    Fig.\ \ref{fLC08hw} as well as the vertical shift for clarity.}
  \label{fLC09nz}
\end{figure}

\vspace{0mm}\paragraph{\bf GRB 091127/SN 2009nz} Figure \ref{fLC09nz}
presents the light curves of the AG in six bands. The
$g\,'r\,'i\,'z\,'\!$ photometry has been image-subtracted to remove
the host-galaxy flux. All are well fitted by a single power law, which
needed a SN component for the $g\,'r\,'i\,'z\,'\!$ bands. No
$K_\text{s}$-band observations were obtained for this event
\citep{Filgas+11b}. The brightest host galaxy allowed by the data was
included in the model for $JH$ at late times (see Table
\ref{tHosts}). The $k$ and $s$ values reflect strong similarities to
SN 1998bw in the $r\,'i\,'\!$ bands. At the redshift of SN 2009nz
($z=0.490$), the $g\,'\!$ band is probing wavelengths centred at
$\sim3000$ \AA, where the flux is strongly affected by absorption-line
blanketing of metals, and so the intensity can differ from SN to
SN. Moreover, since the $U$ band, the bluest band from which the SN
1998bw templates are constructed, is sensitive $\ga3000$ \AA,
extrapolations dominate the $g\,'\!$-band template. Given also the
non-detections after day 12, we derived an upper limit of
$k_{g\,'\!}<1.21$ from the fitting (Table \ref{tSNpars}).

\begin{table*}
  \caption{Parameters of the AG component and goodness of the
    light-curve modelling.}
  \label{tAGpars}
  \centering
  \begin{tabular*}{\textwidth}{@{\extracolsep{\fill}}llllcc}
    \hline\hline\noalign{\vspace{0.5\smallskipamount}}
    GRB &\multicolumn{1}{c}{$\alpha_1$} &\multicolumn{1}{c}{$t_\text{break}$ [days]} &\multicolumn{1}{c}{$\eta$} &$\alpha_2$ &$\chi^2/\mu$ \vspace{0.1\smallskipamount} \\
    \hline\noalign{\vspace{0.5\smallskipamount}}
    081007  &$-0.66\pm0.01$  &$0.91\pm0.05$  &15 fixed     &$-1.40\pm0.05$  &1.5 \\
    091127  &$-0.38\pm0.01$\tablefootmark{a}  &$0.34\pm0.01$\tablefootmark{a}  &$1.3\pm0.1$\tablefootmark{a}  &$-1.63\pm0.02$  &1.4 \\
    101219B &$-1.01\pm0.01$  &$\cdots$        &$\cdots$      &$\cdots$         &1.8  \\
    \hline
  \end{tabular*}
  \tablefoot{The primary power-law slope is $\alpha_1$. In case of a 
    break in the light curve, a secondary slope $\alpha_2$ along with
    the break time $t_\text{break}$ and break smoothness parameter 
    $\eta$ are introduced. The ratio $\chi^2/\mu$ is computed in the
    multiple-component fitting procedure, which includes AG plus 
    SN modelling. See Table \ref{tSNpars} for the SN parameters.
    \tablefoottext{a}{Parameters were taken from the fitting of the 
      full GROND $r\,'\!$-band light curve by \citet{Filgas+11b} 
      except $\alpha_2$, which was fitted by a single power law 
      using the data presented in Fig.\ \ref{fLC09nz} only.}
  }
\end{table*}
%%label:tAGpars

\begin{table*}
  \caption{Parameters of the SN component with respect to SN
    1998bw templates.}
  \label{tSNpars}
  \centering
  \begin{tabular*}{\textwidth}{@{\extracolsep{\fill}}lllclcl}
    \hline\hline\noalign{\vspace{0.5\smallskipamount}}
    SN  &GRB &\multicolumn{1}{c}{Stretch}      &\multicolumn{4}{c}{Luminosity ratio ($k$)\tablefootmark{a}} \\
        &    &\multicolumn{1}{c}{factor ($s$)} &$g\,'$ &\multicolumn{1}{c}{$r\,'$} &$i\,'$ &\multicolumn{1}{c}{$z\,'$} \vspace{0.1\smallskipamount}  \\
    \hline\noalign{\vspace{0.5\smallskipamount}}
    2008hw  &081007 &$0.85\pm0.11$   &$<0.90$ &$0.80\pm0.10$ &$0.65\pm0.08$ &$0.69\pm0.10$ \\
    2009nz  &091127 &$1.03\pm0.04$   &$<1.21$ &$1.15\pm0.09$ &$0.96\pm0.14$ &$0.73\pm0.12$ \\
    2010ma\tablefootmark{b} &101219B &$0.76\pm0.10$   &$0.85\pm0.17$ &$1.78^{\,+\,0.08}_{\,-\,0.17}$ &$1.36\pm0.09$ &$0.63\pm0.09$ \vspace{0.3\smallskipamount} \\
    \hline
  \end{tabular*}
  \tablefoot{
    \tablefoottext{a}{Luminosity ratios are all corrected for Galactic 
      and host-galaxy extinction. The latter correction is taken from 
      the AG SED fitting in Sect.\ \ref{s4_SEDs}.}
    \tablefoottext{b}{No host-galaxy contribution was assumed. See text 
      for estimations including host emission.}
  }
\end{table*}
%%label:tSNpar

\begin{figure}
  \centering
  \includegraphics[bb=4 3 597 584,width=\linewidth,clip]{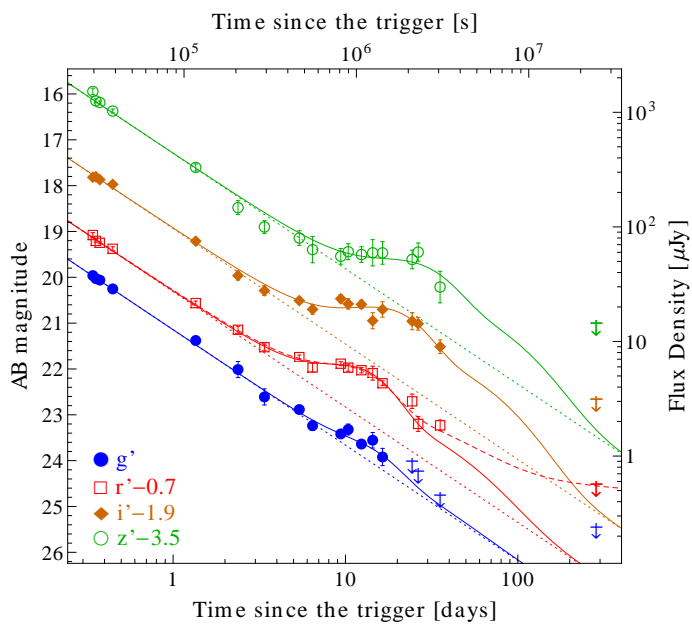}
  \caption{Multicolour GROND light curves of GRB 101219B/SN 2010ma
    corrected for Galactic extinction. The symbol and line coding is
    the same as Fig.\ \ref{fLC08hw} as well as the vertical shift for
    clarity. The red dashed line represents a model with an extra
    host-galaxy component.}
  \label{fLC10ma}
\end{figure}

\vspace{0mm}\paragraph{\bf GRB 101219B/SN 2010ma} Figure \ref{fLC10ma}
shows the GROND light curves of the optical counterpart. The SN bump
is clearly seen in the $r\,'i\,'z\,'\!$ bands, however, it is less
significant in the $g\,'\!$ band. At the redshift of the event, the
$g\,'\!$\ band actually probes the UV regime, therefore, the lower
$g\,'\!$-band SN luminosity is explained by a combination of both the
wavelength extrapolation of the template and the UV line blanketing by
metals. Even though the host galaxy remained undetected, it may
explain the flux excess 35 days after the burst in the $r\,'\!$ band
(dashed line in Fig.\ \ref{fLC10ma}). The $k$ value would decrease
$\sim14$\% in this case. Therefore, the lower error in $k_{r\,'\!}$
was increased to match the 3$\sigma$ lower limit when assuming the
brightest host component possible (Table \ref{tSNpars}). We note that
the $k$ value for the $z\,'\!$\ band is smaller compared to the bluer
bands. Along with the differences in the SN luminosity ratio among all
bands, this indicates that the colours of SN 2010ma are different from
those of SN 1998bw.

\begin{figure}
  \centering
  \includegraphics[bb=29 12 478 460,width=\linewidth,clip]{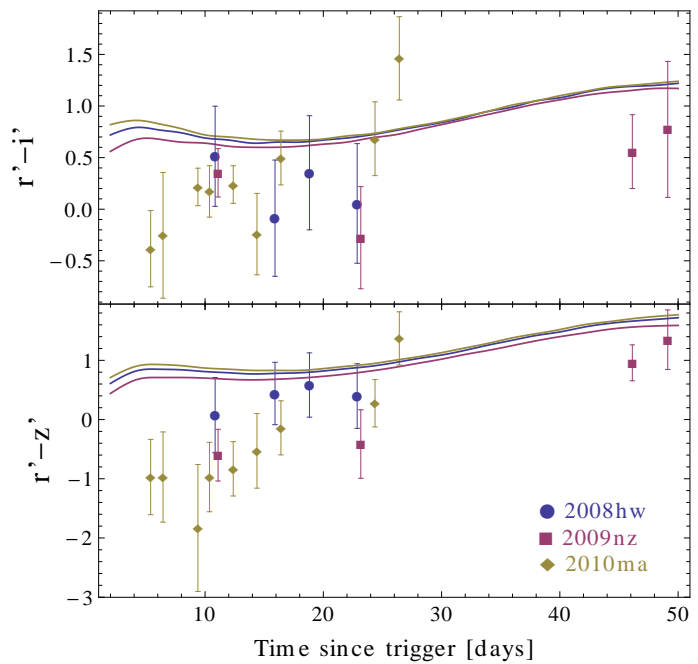}
  \caption{Colour curves corrected for the total extinction of SNe
    2008hw (blue circles), 2009nz (purple squares), and 2010ma (gold
    diamonds) after AG and host subtraction. Blue, purple, and gold
    solid lines are computed from the templates of SN 1998bw at the
    redshifts of SNe 2008hw, 2009nz, and 2010ma, respectively.}
  \label{fColour}
\end{figure}

Fig.\ \ref{fColour} shows the colour curves of the three SNe analysed
compared against the templates of SN 1998bw, where the bluer emission
of SN 2010ma is significant at early times.

\subsection{Spectral energy distributions}\label{s4_SEDs}
Using the available X-ray data from the \emph{Swift}/XRT, the
UV/optical data from the \emph{Swift}/UVOT, and the optical/NIR data
from GROND, we constructed a single AG SED per event with the main
purpose of determining the extinction along the line of sight through
the host galaxy. The SED modelling was performed similarly as in
\citet{Greiner11} and the results are presented in Table
\ref{tSEDpars}.

\begin{table*}
  \caption{Parameters of the SED modelling of the AG.}
  \label{tSEDpars}
  \centering
  \begin{tabular*}{\textwidth}{@{\extracolsep{\fill}} l c c c c l}
    \hline\hline\noalign{\vspace{0.5\smallskipamount}}
    GRB &$\beta_X$ &$A_{V,\text{host}}$ &$E_\text{break}$ &$N_\text{H,host}$ &\multicolumn{1}{c}{$\chi^2/\mu$} \\
    & &[mag] &[eV] &[$10^{21}$\,cm$^{-2}$] & \vspace{0.1\smallskipamount} \\
    \hline\noalign{\vspace{0.5\smallskipamount}}
    081007  &$0.97\pm0.09$ &$0.68\pm0.08$ (SMC) &$37^{\,+\,54}_{\,-\,12}$ &$5.6\pm0.7$ &1.1 \vspace{0.5\smallskipamount} \\
    091127\tablefootmark{a} &$0.748\pm0.004$ &$<0.03$ (LMC) &$2.6\mbox{\,--\,}29.9$ &$0.32\pm0.06$ &$1.1$ \vspace{0.3\smallskipamount} \\
    101219B &$1.12\pm0.01$ &$0.12\pm0.01$ (SMC) &9.0 fixed &$0.6\pm0.3$ &0.8 \\
    \hline
  \end{tabular*}
  \tablefoot{Obeying the fireball model for GRB AGs, the
    high-energy ($\beta_X$) and the low-energy ($\beta_\text{opt}$) 
    spectral indexes are correlated by $\beta_X=\beta_\text{opt}+0.5$ 
    for $E_\text{break}\approx E_\text{cooling}$, where the latter comes 
    from the cooling frequency of the electrons, except in the case of 
    GRB 091127 were $\beta_\text{opt}$ varies in the range 
    $0.25\mbox{\,--\,}0.62$.
    \tablefoottext{a}{The quoted values 
      of $\beta_X$, $N_\text{H,host}$, and reduced $\chi^2$ are computed 
      from the simultaneous best fit to all eight GROND/XRT SEDs by 
      \citet{Filgas+11b}. The $E_\text{break}$ range comes from an 
      observed evolution of $\beta_\text{opt}$. The $A_{V,\text{host}}$ upper 
      limit was taken from \citet{Schady12}.}
  }
\end{table*}
 %%label:tSEDpars

Note that the AG may probe a slightly different line of sight than the
SN photosphere. If anything, the extinction for the SN should be
larger than for the AG, because the AG forms further out, where the
material ejected by the GRB hits the circumstellar medium. In the
standard fireball shock model, this radius is about $10^{17}$ cm
\citep[and even larger for low-luminosity
events;][]{Molinari07}. Moreover, dust can be formed in the SN ejecta,
although not significant amounts on such short timescales
\citep[e.g.,][]{Smith+12}. Therefore, we considered that the
extinction determined through the AG SED is valid for the SN component
as well. The following corresponds to a description of the SED fitting
for each of the rebrightenings.

\begin{figure}
  \centering
  \includegraphics[width=.9\linewidth,clip]{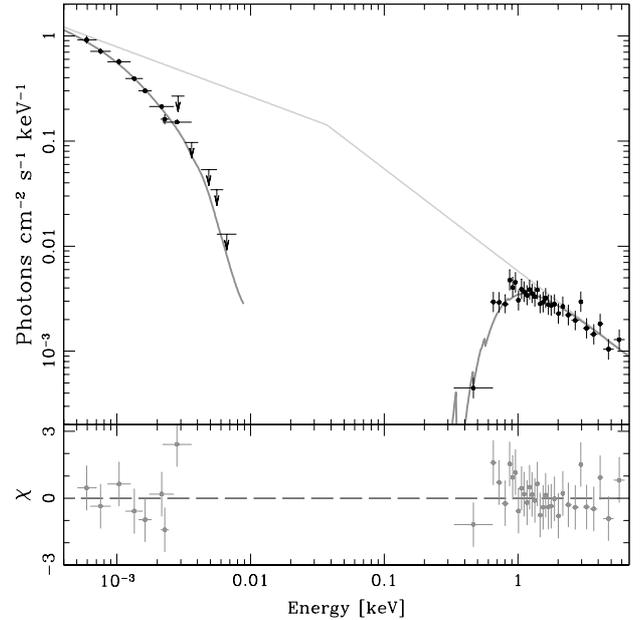}
  \caption{Broad-band AG SED of GRB 081007 at 1.6 ks after
    trigger. The arrows are $3\sigma$ upper limits. The best-fit model
    (thick line) is an extinguished broken power law. The thin line
    represents the unextinguished model. The residuals are in units of
    $\chi$ (lower panel).}
  \label{fBB081007}
\end{figure}

\vspace{0mm}\paragraph{\bf GRB 081007/SN 2008hw} To include
contemporaneous \emph{Swift}/UVOT data, the second GROND epoch was
chosen to study the broad-band SED of GRB 081007 presented in
Fig.\ \ref{fBB081007}. From the UVOT, upper limits in the UV bands are
included, which help to constrain the host-galaxy extinction. The
time-integrated \emph{Swift}/XRT spectrum was interpolated to the
epoch of the UV/optical observations. The resulting values of
host-galaxy extinction and their corresponding statistical uncertainty
are consistent with those computed by \citet{Covino+13} and for the
GROND filters we obtain $A_{g\,'\!,\text{host}}=1.39\pm0.16$,
$A_{r\,'\!,\text{host}}=0.99\pm0.12$,
$A_{i\,'\!,\text{host}}=0.77\pm0.09$,
$A_{z\,'\!,\text{host}}=0.63\pm0.07$, $A_{J,\text{host}}=0.38\pm0.04$,
$A_{H,\text{host}}=0.24\pm0.03$, and $A_{K,\text{host}}=0.14\pm0.02$
in the observer's frame, all in units of magnitude. These values were
used to correct the SN luminosity ratios shown in Table \ref{tSNpars}.

\begin{figure}
  \centering
  \includegraphics[width=.9\linewidth,clip]{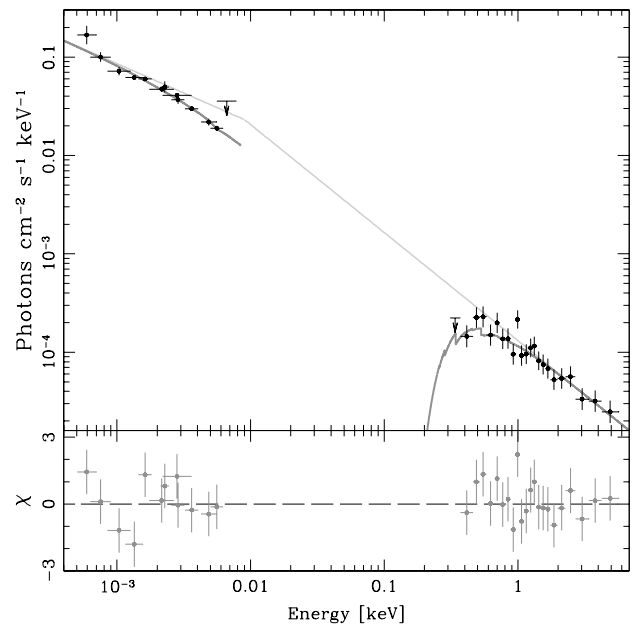}
  \caption{Broad-band AG SED of GRB 101219B at 9.0 h after
    trigger. The symbols, the line coding, and the panels are the same
    as in Fig.\ \ref{fBB081007}.}
  \label{fBB101219B}
\end{figure}

\vspace{0mm}\paragraph{\bf GRB 091127/SN 2009nz} The broad-band SEDs
of the early AG of GRB 091127 were presented by \citet{Filgas+11b}
using the GROND data. A detailed analysis by \citet{Schady12} includes
\emph{Swift}/UVOT and GROND data and constrains the host-galaxy
extinction, which results in $A_{V,\text{host}}<0.03$ mag. The SED
parameters are shown in Table \ref{tSEDpars}.

\vspace{0mm}\paragraph{\bf GRB 101219B/SN 2010ma} Using GROND, XRT,
and UVOT data combined, the AG SED of GRB 101219B was constructed at 9
h after the burst. Figure \ref{fBB101219B} shows a broken power law as
the best fit. The values of the required host-galaxy extinction for
the GROND filters and their corresponding statistical uncertainty in
the observer's frame are $A_{g\,'\!,\text{host}}=0.25\pm0.03$,
$A_{r\,'\!,\text{host}}=0.18\pm0.02$,
$A_{i\,'\!,\text{host}}=0.14\pm0.02$,
$A_{z\,'\!,\text{host}}=0.11\pm0.01$, $A_{J,\text{host}}=0.07\pm0.01$,
$A_{H,\text{host}}=0.04\pm0.01$, and
$A_{K,\text{host}}=0.026\pm0.003$, all in units of magnitude.

\subsection{Calculation and modelling of the bolometric light curves}
\label{s4_Bols}
To isolate the SN from the AG evolution, the light-curve models
computed in Sect.\ \ref{s4_LCs} were employed. The AG contribution was
calculated from the model for the epochs when the SN bump was observed
and it was subtracted from the light curves for each filter. The
uncertainties in the model were appropriately propagated to the final
magnitude errors. After the AG subtraction, quasi-bolometric light
curves were computed for each of the three events by numerically
integrating the monochromatic fluxes in the wavelength range from 340
to 700 nm. The redshift-based luminosity distances in Table
\ref{tSample} were employed to transform observed into absolute
flux. The total uncertainty in the luminosity distance is about 10\%
and has not been included in the quasi-bolometric light curves.

\subsubsection{NIR bolometric correction}\label{s3_NIRcor}
The NIR luminosity proves critical when estimating the bolometric flux
and consequently the physical parameters of the explosion obtained via
the quasi-bolometric flux. However, SNe 2008hw, 2009nz, and 2010ma
remained undetected in the $JHK$ bands. To account for the NIR flux of
these SNe, we proceeded with two different methods. First of all, we
defined the NIR flux from 700 to 2200 nm in the rest frame and the
quasi-bolometric flux from 340 to 2200 nm. The quantity to be
estimated via the two methods is the ratio between the NIR flux and
the quasi-bolometric flux as defined above.

\begin{figure}
  \centering
  \includegraphics[bb=8 3 678 781,width=\linewidth,clip]{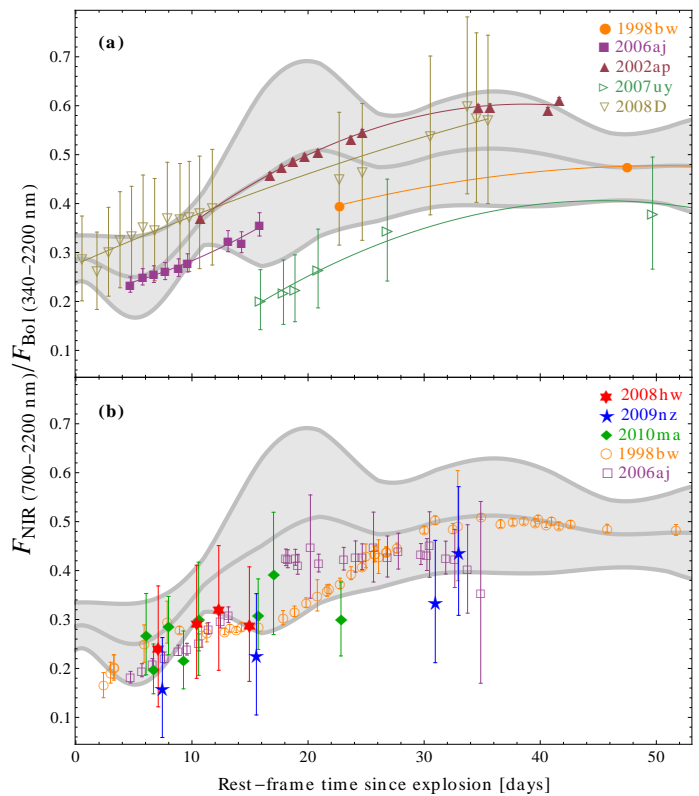}
  \caption{The NIR fraction (700\,--\,2200 nm) of the quasi-bolometric
    flux (340\,--\,2200 nm) {for SE SNe.} {\textit{(a)} The values
      derived using optical/NIR data of five SE SNe are fitted
      separately (coloured solid lines), averaged, and interpolated
      (grey solid line and 1$\sigma$ contours).}  {\textit{(b)} The
      estimations from the BB fits to optical data are shown for five
      GRB-SNe along with the contours from the top panel.} See main
    text for the references of the data sources.}
  \label{fNIRcor}
\end{figure}

The first method consisted in estimating the NIR fraction of the
quasi-bolometric flux using the observed NIR data available in the
literature. {With the optical/NIR} photometry, we computed the ratio
between the NIR and the quasi-bolometric fluxes for the GRB-SNe 1998bw
\citep{Kocevski07} and 2006aj \citep{Patat+01}, and for the type-Ib/c
SNe 2002ap \citep{Foley+03,Yoshii+03}, 2007uy \citep{Roy+13}, and
2008D \citep{Modjaz+09}. {We show the observed NIR fractions in the
  upper panel of Fig.\ \ref{fNIRcor} along with a quadratic-polynomial
  fit for each SN. For each fit, we also obtained the corresponding
  uncertainty as a function of time. By taking the weighted average
  per time bin (5-days width, $t_i=1$ d), we derived the joint
  evolution of the NIR fraction for the five SNe. The non-weighted RMS
  was taken as the 1$\sigma$ error. Only the first time bin uses data
  from a single event and the error here is approximated by the
  uncertainty in the individual polynomial fit. The binned NIR
  fraction was interpolated using quadratic polynomials to retrieve
  values for a given time and to plot the grey contours in
  Fig.\ \ref{fNIRcor}. The SNe 2007uy and 2008D do not contribute much
  to the weighted average, because their host extinction is large and
  highly uncertain \citep[0.3\,--\,0.5 mag;][]{Mazzali+08,Roy+13}.}

{The second method assumes that the SN atmosphere at early phases
  resembles a cooling black body \citep[BB;
    e.g.,][]{Arnett82,Filippenko97,DessartHillier05}. We defined a
  simple BB model with the temperature and a flux normalisation as
  free parameters. Then, we modelled the SN SEDs constructed using
  $g\,'r\,'i\,'z\,'\!$ data at each epoch. The results of the fitting
  procedure are shown in Appendix \ref{app3}. We obtain colour
  temperatures decreasing with time and consistent with other SE SNe
  \citep[e.g.,][]{Folatelli+14}. The extrapolation into the NIR range
  delivered NIR fractions plotted in the lower panel of
  Fig.\ \ref{fNIRcor} with uncertainties between 0.05 and 0.13. The
  increasing NIR flux with time is consistent with the scenario of the
  cooling envelope. We repeated this procedure for the optical data of
  SNe 1998bw \citep[$UBVRI$;][]{Galama98b} and 2006aj
  \citep[$UBVR$;][]{Pian+06,Sollerman06} with results that are
  consistent with those for SNe 2008hw, 2009nz, and 2010ma. Moreover,
  the 1$\sigma$ contours (grey-shaded region) are compatible with the
  all BB estimates derived using optical photometry solely.}

\begin{figure}
  \centering
  \includegraphics[bb=2 2 600 528,width=\linewidth,clip]{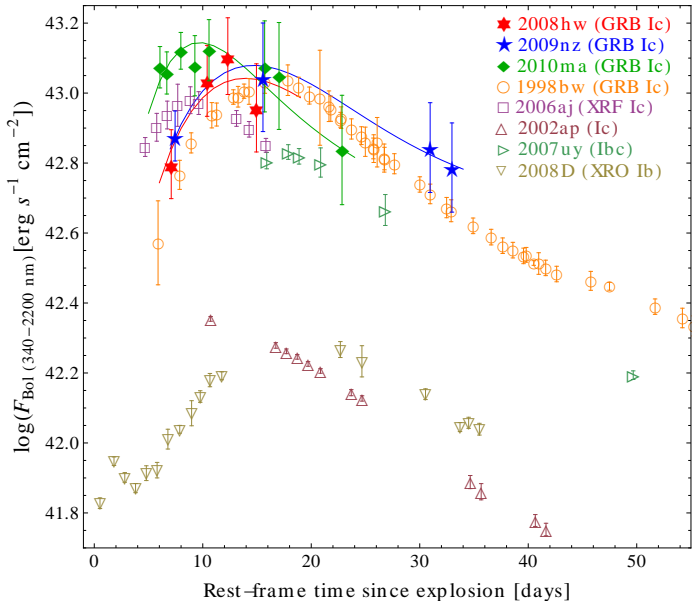}
  \caption{The quasi-bolometric (340\,--\,2200 nm) light curves of SNe
    2008hw (red six-pointed stars), 2009nz (blue five-pointed stars),
    and SN 2010ma (green filled diamonds). The analytic model is shown
    for each SN with the respective colour (details in
    Sect.\ \ref{s4_Bols}). A handful of other type-Ib/c SNe are shown
    for comparison (see main text for the references).}
  \label{fBolLCs}
\end{figure}

{The final NIR correction applied to the optical data was the average
  value between the estimates from the available NIR data for GRB-SNe
  and the estimates from the BB fits. A conservative proxy of the
  NIR-fraction error was chosen to be the largest among the difference
  between the two estimates and their respective errors. Errors
  fluctuate between 0.07 and 0.22.}  We note that the NIR correction
implies $JHK$ magnitudes at maximum consistent with the upper limits
presented in Figs. \ref{fLC08hw}, \ref{fLC09nz}, and
\ref{fLC10ma}. For instance, the brightest magnitudes derived from the
NIR correction are $J=22.6$, $H=23.2$, and $K_\text{s}\approx23.6$ mag
for SN 2010ma. {The corrected} measurements of the quasi-bolometric
flux are presented in Fig.\ \ref{fBolLCs} for the GRB-SNe 2008hw,
2009nz, and 2010ma. For comparison, the quasi-bolometric light curves
(340\,--\,2200 nm) for other SE SNe are also computed and plotted in
Fig.\ \ref{fBolLCs}. We note that all three events lie at luminosity
comparable to that of GRB-SNe 1998bw and 2006aj and are brighter than
``normal'' type-Ib/c SNe. Similar to the results we obtain for
individual optical filters, SN 2010ma turns out to be brighter than SN
1998bw. The quasi-bolometric fluxes of SNe 2008hw, 2009nz, and 2010ma
at maximum {(Table \ref{tPhysPars})} are comparable to
$(1.07\pm0.07)\times10^{43}$ erg s$^{-1}$ for SN 1998bw.

\subsubsection{Physical parameters of the explosion}
The nickel mass $M_\text{Ni}$, the ejected mass $M_\text{ej}$, and the
kinetic energy $E_\text{k}$ of the explosion were extracted from the
luminosity models following the analytic approach by \citet{Arnett82}
for $^{56}$Ni-powered SNe \citep[see,
  e.g.,][]{Maeda03,Taubenberger+06,Valenti+08,Pignata+11,Olivares+12,Roy+13}.
We therefore employed the following expression to model the bolometric
luminosity:

\begin{equation}
  L(t) = M_\text{Ni}\,e^{-x^2} \left[
    (\epsilon_\text{Ni}-\epsilon_\text{Co}) \int_0^x A(z)\,dz +
    \epsilon_\text{Co} \int_0^x B(z)\,dz \right]\text{,}
  \label{eLum}
\end{equation}

\noindent where $A(z)=2z\,e^{-2zy+z^2}$, $B(z)=2z\,e^{-2zy+2zs+z^2}$,
$x\equiv t/\tau_m$, $y\equiv \tau_m/(2\tau_\text{Ni})$, and $s\equiv
\tau_m(\tau_\text{Co}-\tau_\text{Ni})/(2\tau_\text{Co}\tau_\text{Ni})$.
The decay energy for $^{56}$Ni and $^{56}$Co are
$\epsilon_\text{Ni}=3.90\times10^{10}$ erg s$^{-1}$ g$^{-1}$ and
$\epsilon_\text{Co}=6.78\times10^9$ erg s$^{-1}$ g$^{-1}$,
respectively \citep{SutherlandWheeler84,Cappellaro+97}. The decay
times are $\tau_\text{Ni}=8.77$ d and $\tau_\text{Co}=111$ d. The time
scale of the light curve is expressed as

\begin{equation}
  \tau_m=\left(\frac{k_\text{opt}}{\beta c}\right)^{1/2}
  \left(\frac{10M_\text{ej}^3}{3E_\text{k}}\right)^{1/4} \text{,}
  \label{eTau}
\end{equation}

\noindent where $\beta\simeq13.8$ is an integration constant
\citep{Arnett82}, $c$ is the speed of light, and $k_\text{opt}$ is the
optical opacity, {which stays constant in time for this modelling
  scheme. In reality, the opacity depends on the composition and
  temperature of the ejecta, therefore, it changes as the SN
  expands. Assuming a variable opacity, the models by \citet{Chugai00}
  for the bolometric light curve of SN 1998bw deliver an average value
  of 0.07 cm$^2$ s$^{-1}$ for the first 20 days after the explosion.}
{The models by \citet{Mazzali+00} can reproduce the light curve of the
  type-Ic SN 1997ef at early times using a constant opacity of 0.08
  cm$^2$ s$^{-1}$. With a constant opacity of 0.06 cm$^2$ s$^{-1}$,
  the synthetic light curves by \citet{Maeda03} manage to reproduce
  the data of hypernovae at early phases. Thus, we chose a value of
  $k_\text{opt}=0.07\pm0.01$ cm$^2$ s$^{-1}$, which includes within
  1$\sigma$ the opacity values that have been employed in the
  literature.}

Equations \ref{eLum} and \ref{eTau} are valid only for the
photospheric phase ($t-t_0\la40$ d). Given the lack of detections
beyond day 40, no nebular component has been considered \citep[see
  appendix in][for the complete model]{Valenti+08}. The modelling
procedure employed consists of a weighted $\chi^2$ minimisation, where
$M_\text{Ni}$ and $M_\text{ej}^{\,3}/E_\text{k}$ are free. The latter
will be dubbed the ``timescale parameter'' hereafter, because it
approximates the light-curve shape (see Sect.\ \ref{c4_Diss} for
details). To compute $M_\text{ej}$ and $E_\text{k}$ from the timescale
parameter, we used the expression for the photospheric expansion
velocity at maximum luminosity from \citet{Arnett82}:

\vspace{-5mm}
\begin{equation}
  \upsilon_\text{ph}^{\,2}\approx\frac{3}{5}\,\frac{2E_\text{k}}{M_\text{ej}}\text{.}
  \label{eVph}
\end{equation}
\vspace{-4mm}

\noindent This quantity is critical to obtain reliable physical
parameters of the explosion \citep{Mazzali+13}. A minimum expansion
velocity of $\sim$ 14,000 km s$^{-1}$ \citep[SN 1998bw;][]{Pian+06}
and a maximum of $\sim$ 28,000 km s$^{-1}$ \citep[SN
  2010bh;][]{Bufano+12} {have been measured for GRB-SNe. Thus, we
  employed $22,000\pm4,000$ km s$^{-1}$} if estimates of the
photospheric velocity are not available. This {conservative proxy
  encompasses with a 2$\sigma$ confidence} the photospheric velocity
of most spectroscopically-confirmed GRB-SNe at maximum luminosity
\citep[see, e.g.,][]{Bufano+12}.

To calculate uncertainties, we performed thousand Monte-Carlo
simulations for each event. {Assuming Gaussian errors,} each
simulation consisted of a $\chi^2$ minimisation {between the model
  with a randomised opacity and the randomised quasi-bolometric data
  points.}  From the resulting distributions for $M_\text{Ni}$ and
$M_\text{ej}^3/E_\text{k}$, we obtain the median and the standard
deviation (1$\sigma$). Then, eq.\ \ref{eVph} is employed to compute
$M_\text{ej}$ and $E_\text{k}$ propagating the errors accordingly.
When using the wide range of expansion velocities for SNe 2008hw and
2010ma, we computed the weighted average of $M_\text{ej}$ and
$E_\text{k}$ between the parameters obtain using the minimum and
maximum photospheric velocities as defined above, and listed the
corresponding 1$\sigma$ ranges in Table \ref{tPhysPars}. For SN
2009nz, \citet{Berger+11} measure an expansion velocity of 17,000 km
s$^{-1}$ from \ion{Si}{ii} $\lambda$6355, which has been identified as
a reliable tracer of the photospheric velocity
\citep{Sauer+06,Valenti+08}. Although the date of the spectrum (16.3
rest-frame days after the GRB) coincides quite well with the maximum
luminosity, the spectral coverage barely extends to 6250 \AA\ and the
spectrum has low S/N. Therefore, we assigned to this velocity a
conservative uncertainty of {1,500} km s$^{-1}$, which corresponds to
about {30} \AA.  The physical parameters and best-fit models are
listed and plotted in Table \ref{tPhysPars} and Fig.\ \ref{fBolLCs},
respectively.

\begin{table*}
  \caption{Physical parameters from quasi-bolometric light curves.}
  \label{tPhysPars}
  \centering
  \begin{tabular*}{\textwidth}{@{\extracolsep{\fill}} l l c c c c c c c }
    SN &\multicolumn{1}{c}{$M_\text{Ni}$}       &$M_\text{ej}$        &$E_\text{k}$  &$M_\text{ej}^{\,3}/E_\text{k}$  &$\upsilon_\text{ph}$ &$\log{L_\text{max}}$ &$t_\text{max}$ &Reference \vspace{0.1\smallskipamount} \\
    \hline\hline\noalign{\vspace{0.5\smallskipamount}}
       &\multicolumn{1}{c}{$[\text{M}_\sun]$}  &$[\text{M}_\sun]$   &$[10^{51}\,\text{erg}]$ &$[10^{-51}\,\text{M}_\sun^{\,3}\,\text{erg}^{-1}]$ &[$10^3$ km s$^{-1}$] &[erg s$^{-1}$] &[days]  \vspace{0.1\smallskipamount} \\
    \hline\noalign{\vspace{0.5\smallskipamount}}
    2008hw  &$0.39^{\,+\,0.08}_{\,-\,0.04}$       &$2.3^{\,+\,1.0}_{\,-\,0.7}$     &$19\pm15$           &$0.7^{\,+\,0.7}_{\,-\,0.2}$      &$22\pm4$  &$43.1\pm0.3$ &$12\pm3$ &1   \vspace{0.5\smallskipamount}\\
    2009nz  &$0.50\pm0.04$                   &$2.4^{\,+\,0.6}_{\,-\,0.3}$     &$11\pm4$            &$1.2^{\,+\,0.6}_{\,-\,0.2}$      &$17\pm1.5$ &$43.0\pm0.2$ &$18\pm4$ &1   \vspace{0.5\smallskipamount}\\
    2010ma  &$0.43^{\,+\,0.03}_{\,-\,0.02}$       &$1.3^{\,+\,0.4}_{\,-\,0.3}$     &$10\pm6$            &$0.20^{\,+\,0.09}_{\,-\,0.04}$    &$22\pm4$  &$43.1\pm0.2$ &$10\pm2$ &1  \vspace{0.5\smallskipamount}\\
    \hline\noalign{\vspace{0.5\smallskipamount}}
\multirow{2}{*}{1998bw}      &0.38\,--\,0.48            &11                         &50                       &27            &\multirow{2}{*}{$17\pm1$}  &\multirow{2}{*}{$43.03\pm0.01$}  &\multirow{2}{*}{17.8}  &2,3\vspace{-0.5\smallskipamount} \\
                             &$0.45\pm0.01$             &$3.4\pm0.2$                &$16\pm3$                 &$2.4\pm0.2$   &                     &                        &                       &1  \vspace{0.5\smallskipamount}  \\
    2003dh                   &0.25\,--\,0.45            &8                          &40                       &13            &$\sim20$  &$\sim43.0$ &$\sim18$ &4  \vspace{0.5\smallskipamount}  \\
    2003lw                   &0.45\,--\,0.65            &13                         &60                       &37            &$\sim18$  &$\sim43.2$ &$\sim18$ &5  \vspace{0.5\smallskipamount}  \\
\multirow{2}{*}{2006aj}      &\multicolumn{1}{c}{0.21}  &2                          &2                        &4.0           &\multirow{2}{*}{$19\pm1$}  &\multirow{2}{*}{$42.99\pm0.02$}  &\multirow{2}{*}{8.8}   &6,7\vspace{-0.5\smallskipamount} \\
                             &$0.26\pm0.01$             &$0.6\pm0.04$               &$3.5\pm0.6$              &$0.057\pm0.004$ &                     &                        &                       &1  \vspace{0.5\smallskipamount}  \\
    2010bh                   &$0.21\pm0.03$             &$2.6\pm0.2$                &$24\pm7$                 &$0.7\pm0.3$   &28 &42.63 &8.0       &8,9    \\
    \hline\noalign{\vspace{0.5\smallskipamount}}
    2002ap                   &\multicolumn{1}{c}{0.10}  &2.5                        &4                        &3.9           &14   &$\sim42.4$ &$\sim11$  &10   \\
    2003jd                   &\multicolumn{1}{c}{0.36}  &3                          &7                        &3.9           &13.5 &$\sim42.9$ &$\sim17$  &11   \\
    2007uy                   &$0.30\pm0.01$             &$4.4\pm0.3$                &$15\pm1$                 &$5.6\pm1.0$   &15.2 &42.83 &17.9           &12   \\
    2008D                    &0.07\,--\,0.09            &$5.3\pm1.0$                &\ \,$6\pm3$              &$25\pm17$     &$\sim10$ &$\sim42.4$ &$\sim19$      &13,14\\
    2009bb                   &$0.22\pm0.06$             &$4.1\pm1.9$                &$18\pm7$                 &$3.8\pm5.5$   &$\sim20$ &$\sim42.8$ &$\sim18$               &15   \\
    \hline
  \end{tabular*}
  \tablefoot{Different parts of the table correspond to GRB-SNe
    analysed here (top), further GRB-SNe (middle), and other SE SNe
    (bottom). Uncertainties are given at the 1$\sigma$ level. The
    $\upsilon_\text{ph}$ values correspond to measurements as defined
    by eq.\ \ref{eVph}. The $L_\text{max}$ and $t_\text{max}$ values
    are estimated at the maximum bolometric luminosity, where
    $t_\text{max}$ is defined with respect to the explosion time.}
  \tablebib{(1) This paper, (2) \citealt{Iwamoto+98}, (3)
    \citealt{Mazzali+01}, (4) \citealt{Mazzali+03}, (5)
    \citealt{Mazzali+06a}, (6) \citealt{Pian+06}, (7)
    \citealt{Mazzali+06b}, (8) \citealt{Olivares+12}, (9)
    \citealt{Bufano+12}, (10) \citealt{Mazzali+07}, (11)
    \citealt{Valenti+08}, (12) \citealt{Roy+13}, (13)
    \citealt{Mazzali+08}, (14) \citealt{Tanaka+09}, (15)
    \citealt{Pignata+11}.}
\end{table*}

%%label:tPhysPars

Figure\ \ref{fBolLCs} shows that the light curves are reasonably well
modelled within the errors. In the case of SN 2009nz, however,
\citet{Berger+11} obtain a lower $M_\text{Ni}$ ($0.35\,\text{M}_\sun$
of $^{56}$Ni). They scale the $I$-band photometry from
\citet{Cobb+10b} to obtain a $V$-band absolute magnitude at maximum
and then they compute $M_\text{Ni}$ using a simplification of the
formalism by \citet{Arnett82}, which should deliver results similar to
ours. Given that a higher $^{56}$Ni mass implies higher luminosity
\citep{Colgate+80,Arnett82} and our data includes more flux (three
GROND bands plus the NIR correction), our value is likely more
reliable. The $M_\text{ej}$ quantity is consistent with that presented
by \citet{Berger+11}. Regarding SNe 2008hw and 2009nz, no detailed
photometric studies have been published for these yet.

\section{Discussion}\label{c4_Diss}

Regarding the NIR correction utilised in Sect.\ 3, we have to address
that the extrapolation to SNe with different properties is the major
source of uncertainty for this correction, although the five SE SNe
selected for the analysis already cover a wide range of properties. A
clear case of deviation from our NIR correction is the contribution
shown by SN 2002ap. Even though SN 2002ap was not preceded by a GRB,
it was a type-Ic event that showed a maximum NIR {fraction} of about
{0.6} of the total quasi-bolometric flux. Moreover, the colours of SN
2010ma turned out to be significantly bluer before $t<20$ d than those
of SN 1998bw. This could hint at a higher temperature of the SN
envelope and, therefore, lower NIR fluxes. Therefore, we caution that
there might be GRB-SNe that will not fit into our estimation of the
NIR correction.  {The NIR fraction could have variations as large as
  $\pm0.15$ if we compare SN 2002ap to SN 2007uy. This would translate
  into a maximum variation of about $\pm30$\% in the quasi-bolometric
  flux (equivalent to a 1$\sigma$ error of $\sim10$\%)} and therefore
in the determinations of $M_\text{Ni}$. This issue could be solved in
the future by using a larger sample, i.e., by including observations
of new GRB-SNe in the NIR bands.

\begin{figure}
  \centering
  \includegraphics[bb=4 1 699 505,width=\linewidth,clip]{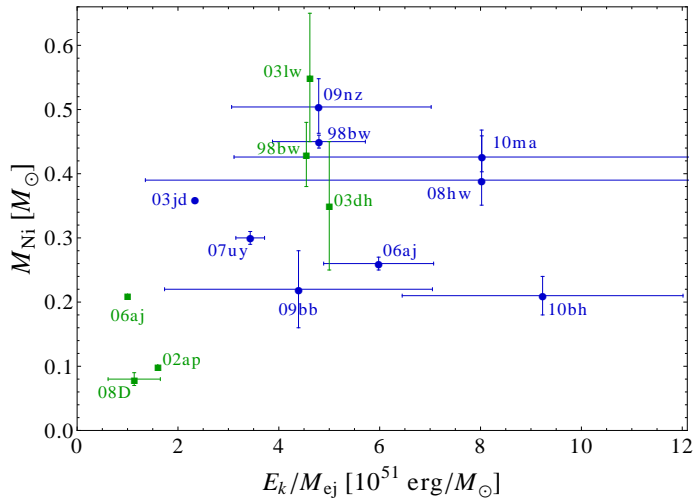}
  \caption{The nickel mass against the envelope energy per unit
    mass. While green squares depict values obtained via
    hydrodynamical simulations, blue circles correspond to parameters
    measured using the analytic approach explained in
    Sect.\ \ref{s4_Bols} or similar. When ranges are given in Table
    \ref{tPhysPars}, we plotted the weighted centre of the range. {We
      caution that no uncertainties are available in the literature
      for some measurements (see Table \ref{tPhysPars}).}}
  \label{fPhys}
\end{figure}

With the purpose of comparing the physical parameters computed by
others for a set of different SNe, we gathered results from the
literature in Table \ref{tPhysPars}, although uncertainties were
unfortunately not available for all. To compare the analytic method
against the hydrodynamical simulations, we additionally computed the
physical parameters of the explosion for SNe 1998bw and 2006aj using
$\upsilon_\text{ph}=17,000$ and 19,000 km s$^{-1}$ \citep{Pian+06},
respectively. In Figure \ref{fPhys} we plotted the kinetic energy per
unit mass $E_\text{k}/M_\text{ej}$ against the synthesised nickel mass
$M_\text{Ni}$, a diagram that have been presented by
\citet{Bufano+12}. Even though some values have large uncertainties,
we recognised a trend where the more energetic the SNe is, the more
$^{56}$Ni it synthesises {\citep{Mazzali+07}}. We note also that
hydrodynamical (green) and analytic (blue) measurements are consistent
for SN 1998bw, despite showing significant differences for
$M_\text{ej}$ and $E_\text{k}$ individually. This is because the ratio
$E_\text{k}/M_\text{ej}$ is proportional to $\upsilon_\text{ph}^{\,2}$
(eq.\ \ref{eVph}), which is a common measurement for both
approaches. The discrepancies in $M_\text{ej}$ and $E_\text{k}$
individually are probably attributed to the different values used for
the optical opacity and of course to the different {assumptions} and
models employed {(hydrodynamical or analytic)}. This would explain the
large discrepancies shown for SN 2006aj as well, where the difference
in $M_\text{Ni}$ could be explained by our inclusion of the NIR
data. We caution that the physical parameters of the explosion might
be highly model-dependant, especially the values obtained for
$M_\text{ej}$ and $E_\text{k}$.

\section{Summary and conclusions}\label{Summary}

Here we studied the GRB-SN connection by means of three individual
events followed up in depth by XRT, UVOT, and GROND. The X-ray, UV,
optical, and NIR data covered approximately six orders of magnitude in
the radiative energy domain. Excluding $\gamma$-ray data, this
represents a very comprehensive data set presented for the
associations GRB 081007/SN 2008hw, GRB 091127/SN 2009nz, and GRB
101219B/SN 2010ma.

In Sect.\ \ref{c3_4Reb}, the light curves of the three events are
thoroughly analysed. The host-galaxy extinction along the line of
sight of each event is computed from the broad-band SED. The light
curves of individual filter bands are modelled with SN 1998bw
templates (Sect.\ \ref{s4_LCs}). The AG component is subtracted to
isolate the SN counterpart. {The NIR flux was estimated from the data
  of five SE SNe and using BB fits of the optical data. This
  correction has been applied} to the integrated optical flux of our
rebrightenings to obtain quasi-bolometric light curves from 340 to
2200 nm. We note that the NIR contribution of SN 2002ap is
10\,--\,15\% larger than that of the {GRB-SNe 1998bw and
  2006aj}. Moreover, the colours of SN 2010ma at early times are bluer
than those of SN 1998bw suggesting lower NIR fluxes for this
object. Therefore, we conclude that more NIR data is needed to
constrain better the NIR contribution in GRB-SN light curves. Using an
analytic model for bolometric light curves, the physical parameters of
the SN explosion were computed for each event analogous to the case of
SN 2010bh in \citet{Olivares+12}. We derived nickel and ejected masses
of about $0.4\text{\,--\,}0.5\,\text{M}_\sun$ and
$1\text{\,--\,}3\,\text{M}_\sun$, respectively, and kinetic energies
of about $10^{52}$ erg, which are higher than those of local type-Ic
SNe and comparable to other GRB-SN events (see Table \ref{tPhysPars}
and Fig.\ \ref{fPhys}).

In conclusion, all three cases exhibit similarities to other GRB-SNe
in terms of luminosity and physical parameters. SN 2008hw turned out
to be somewhat fainter and slightly bluer than SN 1998bw (see Table
\ref{tSNpars}). Moreover, SN 2009nz showed the most similarities with
SN 1998bw in luminosity and evolution. SN 2010ma was significantly
bluer and brighter than SN 1998bw. Both the latter and SN 2010bh have
among the earliest optical peaks ever recorded (approximately 8 days
after the GRB) and fade more rapidly than almost every other GRB-SN,
HN, or typical type-Ic SN.

\begin{acknowledgements}

  We acknowledge the referee for suggestions and corrections that
  helped improving the paper significantly. F.O.E.\ thanks F.~Bufano
  for the sanity checks on the bolometric LCs. The Ph.~D.\ studies of
  F.O.E.\ were funded both by the Deut\-scher Aka\-de\-mi\-scher
  Aus\-tausch Dienst (DAAD) and the Co\-mi\-si\'on Na\-cion\-al de
  In\-ves\-ti\-ga\-ci\'on Cien\-t\'{\i}\-fi\-ca y
  Tec\-no\-l\'o\-gi\-ca (CONICYT). F.O.E.\ acknowledges support from
  FONDECYT through postdoctoral grant 3140326. F.O.E.\ and
  G.P.\ acknowledge support from project IC120009 ``Millennium
  Institute of Astrophysics (MAS)'' of the Iniciativa Cient\'{\i}fica
  Milenio del Ministerio de Econom\'{\i}a, Fomento y Turismo de
  Chile. Part of the GROND funding (both hardware and personnel) was
  generously granted from the Leibniz-Prize to Prof.~G.\ Hasinger,
  Deut\-sche For\-schungs\-ge\-mein\-schaft (DFG) grant HA
  1850/28--1. S.K., A.R., A.N., D.A.K.\ acknowledge support by DFG
  grant KL 766/16-1. S.~S.\ acknowledges support by the Th\"uringer
  Ministerium f\"ur Bildung, Wissenschaft und Kultur under FKZ
  12010-514. D.A.K.\ acknowledges financial support from MPE and TLS.
  A.R., A.N., D.A.K.\ are grateful for travel funding support through
  MPE. This work made use of data supplied by the UK \emph{Swift}
  Science Data Centre at the University of Leicester and data from the
  NASA's Astrophysics Data System (NAS 5--26555). The Dark Cosmology
  Centre is funded by the Danish National Research Foundation.

\end{acknowledgements}

\bibliographystyle{aa}

%\bibliography{../zRef} % your references Yourfile.bib 

\Online
\begin{appendix}
  \section[Optical/NIR photometry]{Optical/NIR photometry}\label{app1}

  \begin{minipage}{\textwidth}
    The three Tables presented as follows are corrected for Galactic
    foreground extinction \citep{Schlegel+98}.
  \end{minipage}

  \begin{table}[h]
  \caption{GRB 081007/SN 2008hw.}
  \label{t081007}      
  \centering
  \begin{minipage}{\textwidth}
%   \begin{tabular*}{1.2\textwidth}{@{\extracolsep{\fill}}cccccc|ccccc}
    \begin{tabular*}{\textwidth}{@{\extracolsep{\fill}}cccccc|ccccc}
      \hline\hline\noalign{\vspace{0.5\smallskipamount}}
      $t-t_0$ &$\Delta t$\,\tablefootmark{a}  &$g\,'$         &$r\,'$         &$i\,'$         &$z\,'$          &$t-t_0$ &$\Delta t$  &$J$     &$H$   &$K_\text{s}$ \\
      {[d]}   &[ks]  &               &               &               &                &[d]    &[ks]   &        &     &  \vspace{0.5\smallskipamount} \\
      \hline\noalign{\vspace{0.5\smallskipamount}}
      0.013 &0.39 &$17.25(16)$ &$17.28(11)$ &$17.14(09)$ &$17.02(08)$ &0.013 &0.40 &$17.22(05)$ &$16.76(03)$ &$16.56(05)$ \\
      0.018 &0.40 &$17.51(16)$ &$17.53(11)$ &$17.37(09)$ &$17.29(08)$ &0.019 &0.42 &$17.53(05)$ &$17.06(03)$ &$16.82(04)$ \\
      0.025 &0.69 &$17.77(16)$ &$17.78(11)$ &$17.64(09)$ &$17.53(07)$ &0.026 &0.75 &$17.77(05)$ &$17.28(03)$ &$17.05(04)$ \\
      0.035 &0.70 &$17.99(16)$ &$17.92(11)$ &$17.85(09)$ &$17.76(08)$ &0.035 &0.76 &$17.98(05)$ &$17.44(05)$ &$17.26(04)$ \\
      0.050 &1.74 &$18.28(16)$ &$18.18(11)$ &$18.11(09)$ &$17.96(07)$ &0.050 &1.79 &$18.16(05)$ &$17.70(03)$ &$17.51(04)$ \\
      0.071 &1.75 &$18.51(16)$ &$18.48(11)$ &$18.34(09)$ &$18.25(07)$ &0.071 &1.80 &$18.39(05)$ &$17.94(04)$ &$17.74(04)$ \\
      0.092 &1.72 &$18.68(16)$ &$18.67(11)$ &$18.57(09)$ &$18.50(07)$ &0.092 &1.77 &$18.58(05)$ &$18.10(04)$ &$17.90(05)$ \\
      0.912 &2.53 &$20.42(16)$ &$20.29(13)$ &$20.15(10)$ &$20.03(08)$ &0.902 &0.75 &$20.11(07)$ &$ >19.41  $ &$ >19.16  $ \\
      3.807 &1.73 &$22.00(34)$ &$22.09(27)$ &$21.99(28)$ &$21.88(28)$ &0.917 &1.79 &$20.20(06)$ &$19.59(10)$ &$19.54(10)$ \\
      10.85 &5.33 &$ >23.46  $ &$23.32(15)$ &$22.99(23)$ &$23.17(25)$ &3.808 &1.78 &$ >20.96  $ &$ >20.03  $ &$ >19.24  $ \\
      15.91 &5.32 &$ >23.87  $ &$23.17(20)$ &$23.20(35)$ &$ >22.43  $ &10.85 &5.37 &$ >21.67  $ &$ >20.84  $ &$ >19.44  $ \\
      18.83 &7.17 &$ >23.75  $ &$23.27(30)$ &$22.97(32)$ &$22.76(31)$ &15.91 &5.38 &$ >21.67  $ &$ >20.76  $ &$ >19.83  $ \\
      22.86 &7.11 &$ >24.03  $ &$23.51(25)$ &$23.45(38)$ &$23.15(36)$ &18.83 &7.21 &$ >21.15  $ &$ >19.67  $ &$ >19.62  $ \\
      28.89 &3.53 &$ >23.29  $ &$ >22.78  $ &$ >22.18  $ &$ >21.12  $ &22.86 &7.16 &$ >21.43  $ &$ >20.69  $ &$ >20.00  $ \\
      \hline                  
    \end{tabular*}
    \tablefoot{The GRB trigger time is $t_0=54746.225$ MJD. All data
      corrected for $A_{V,\text{host}}=0.68\pm0.08$ mag (SMC). Image
      subtraction of the host was performed for $g\,'r\,'i\,'z\,'\!$.
      \tablefoottext{a}{The duration of the observation.}  }
  \end{minipage}
\end{table}

  \begin{table}[h]
  \caption{GRB 091127/SN 2009nz.}
  \label{t091127}      
  \centering
  \begin{minipage}{\textwidth}
%   \begin{tabular*}{1.1\textwidth}{@{\extracolsep{\fill}}cccccc|ccccc}
    \begin{tabular*}{\textwidth}{@{\extracolsep{\fill}}cccccc|ccccc}
      \hline\hline\noalign{\vspace{0.5\smallskipamount}}
      $t-t_0$ &$\Delta t$\,\tablefootmark{a}  &$g\,'$         &$r\,'$         &$i\,'$         &$z\,'$          &$t-t_0$ &$\Delta t$  &$J$     &$H$  \\
      {[d]}   &[ks]  &               &               &               &                &[d]    &[ks]   &        & \vspace{0.5\smallskipamount} \\
      \hline\noalign{\vspace{0.5\smallskipamount}}
      1.075 &1.55 &$19.69(03)$ &$19.50(05)$ &$19.37(03)$ &$19.17(01)$ &1.070 &0.75 &$19.04(08)$ &$18.81(09)$ \\
      $\cdots$ &$\cdots$ &$\cdots$ &$\cdots$ &$\cdots$ &$\cdots$ &1.080 &0.73 &$19.09(08)$ &$18.88(09)$ \\
      1.260 &0.69 &$19.94(03)$ &$19.77(03)$ &$19.69(03)$ &$19.46(03)$ &1.243 &1.75 &$19.23(06)$ &$18.99(07)$ \\
      2.079 &1.70 &$20.81(04)$ &$20.58(04)$ &$20.47(01)$ &$20.38(02)$ &2.080 &1.75 &$20.09(09)$ &$19.86(10)$ \\
      2.198 &1.71 &$20.88(03)$ &$20.65(04)$ &$20.54(03)$ &$20.41(03)$ &2.198 &1.77 &$20.09(07)$ &$19.90(12)$ \\
      3.207 &1.71 &$21.64(07)$ &$21.32(04)$ &$21.26(05)$ &$21.13(05)$ &3.207 &1.76 &$20.54(32)$ &$20.52(22)$ \\
      4.212 &1.70 &$22.10(09)$ &$21.87(06)$ &$21.85(10)$ &$21.60(07)$ &4.212 &1.75 &$20.88(18)$ &$20.99(28)$ \\
      6.175 &1.71 &$22.68(13)$ &$22.40(10)$ &$22.51(14)$ &$22.22(10)$ &6.175 &1.75 &$21.53(31)$ &$ >21.26  $ \\
      11.10 &1.71 &$23.76(12)$ &$22.76(06)$ &$22.50(10)$ &$22.86(13)$ &11.10 &1.75 &$ >21.63  $ &$ >21.16  $ \\
      23.17 &0.34 &$ >22.87  $ &$22.85(23)$ &$ >22.05  $ &$ >22.15  $ &$\cdots$ &$\cdots$ &$\cdots$ &$\cdots$ \\
      46.13 &3.92 &$ >24.43  $ &$24.08(13)$ &$23.58(27)$ &$23.20(22)$ &46.13 &3.97 &$21.74(27)$ &$ >21.35  $ \\
      49.12 &1.70 &$ >24.69  $ &$24.49(32)$ &$ >23.41  $ &$23.28(26)$ &49.12 &1.75 &$ >21.69  $ &$ >21.19  $ \\
      54.09 &1.90 &$ >21.26  $ &$ >23.88  $ &$ >18.86  $ &$ >22.51  $ &54.09 &1.75 &$ >21.37  $ &$ >20.93  $ \\
      \hline                  
    \end{tabular*}
    \tablefoot{The GRB trigger time is $t_0=55162.976$ MJD. Data not
      corrected for negligible $A_{V,\text{host}}<0.03$ mag
      (LMC). Image subtraction of the host was performed for
      $g\,'r\,'i\,'z\,'\!$.  \tablefoottext{a}{The duration of the
        observation.}  }
  \end{minipage}
\end{table}

  \clearpage
  \begin{table}[t]
  \caption{GRB 101219B/SN 2010ma.}
  \label{t101219B}
  \centering
  \begin{minipage}{\textwidth}
%   \begin{tabular*}{1.2\textwidth}{@{\extracolsep{\fill}}cccccc|ccccc}
    \begin{tabular*}{\textwidth}{@{\extracolsep{\fill}}cccccc|ccccc}
      \hline\hline\noalign{\vspace{0.5\smallskipamount}}
      $t-t_0$ &$\Delta t$\,\tablefootmark{a}  &$g\,'$         &$r\,'$         &$i\,'$         &$z\,'$          &$t-t_0$ &$\Delta t$  &$J$     &$H$   &$K_\text{s}$ \\
      {[d]}   &[ks]  &               &               &               &                &[d]    &[ks]   &        &     & \vspace{0.5\smallskipamount} \\
      \hline\noalign{\vspace{0.5\smallskipamount}}
      0.342 &0.91 &$19.71(05)$ &$19.60(05)$ &$19.58(06)$ &$19.33(07)$ &0.342 &0.92 &$19.10(09)$ &$19.02(12)$ &$18.73(23)$ \\
      0.356 &1.44 &$19.78(05)$ &$19.73(03)$ &$19.57(03)$ &$19.54(04)$ &0.357 &1.49 &$19.26(08)$ &$19.08(11)$ &$18.84(20)$ \\
      0.376 &1.71 &$19.81(06)$ &$19.77(02)$ &$19.63(03)$ &$19.57(03)$ &0.376 &1.76 &$19.33(07)$ &$18.98(10)$ &$19.07(26)$ \\
      0.446 &1.72 &$20.00(06)$ &$19.90(04)$ &$19.73(03)$ &$19.76(04)$ &0.447 &1.77 &$19.61(10)$ &$19.31(12)$ &$18.74(23)$ \\
      1.353 &2.92 &$21.13(09)$ &$21.09(04)$ &$20.97(05)$ &$20.99(07)$ &1.354 &2.98 &$21.06(22)$ &$20.64(21)$ &$19.88(37)$ \\
      2.381 &1.47 &$21.76(17)$ &$21.67(07)$ &$21.72(09)$ &$21.87(15)$ &6.442 &4.20 &$21.44(39)$ &$21.06(44)$ &$ >19.60  $ \\
      2.508 &1.61 &$ >21.86  $ &$ >22.30  $ &$ >22.02  $ &$ >21.83  $ &2.381 &1.52 &$ >20.95  $ &$21.05(30)$ &$ >19.11  $ \\
      3.387 &1.72 &$22.36(17)$ &$22.05(07)$ &$22.05(10)$ &$22.29(13)$ &2.507 &1.52 &$ >20.43  $ &$ >19.96  $ &$ >19.81  $ \\
      5.399 &3.53 &$22.64(05)$ &$22.26(05)$ &$22.27(06)$ &$22.53(16)$ &3.388 &1.77 &$ >21.10  $ &$ >20.80  $ &$ >20.04  $ \\
      6.445 &3.53 &$22.99(10)$ &$22.49(09)$ &$22.46(10)$ &$22.78(28)$ &5.399 &3.58 &$ >21.38  $ &$ >20.74  $ &$ >19.71  $ \\
      9.395 &6.69 &$23.17(06)$ &$22.41(04)$ &$22.23(08)$ &$22.93(17)$ &9.377 &3.58 &$ >21.10  $ &$ >20.77  $ &$ >19.98  $ \\
      10.38 &3.52 &$23.07(12)$ &$22.49(07)$ &$22.34(11)$ &$22.83(17)$ &10.38 &3.56 &$ >21.37  $ &$ >21.03  $ &$ >20.21  $ \\
      12.38 &3.52 &$23.39(07)$ &$22.55(06)$ &$22.35(09)$ &$22.89(17)$ &12.38 &3.57 &$ >21.30  $ &$ >20.79  $ &$ >20.17  $ \\
      14.39 &3.52 &$23.31(17)$ &$22.61(15)$ &$22.71(17)$ &$22.85(28)$ &14.39 &3.57 &$ >20.67  $ &$ >20.34  $ &$ >19.13  $ \\
      16.41 &6.71 &$23.67(19)$ &$22.83(05)$ &$22.46(16)$ &$22.85(24)$ &16.40 &3.59 &$ >20.95  $ &$ >20.38  $ &$ \cdots $  \\
      24.37 &4.18 &$ >23.72  $ &$23.23(15)$ &$22.72(18)$ &$23.00(20)$ &24.37 &4.23 &$ >21.28  $ &$ >20.67  $ &$ >19.35  $ \\
      26.35 &0.78 &$ >22.36  $ &$ >22.79  $ &$ >21.57  $ &$ >21.59  $ &26.35 &0.83 &$ >20.35  $ &$ >20.11  $ &$ >19.50  $ \\
      26.43 &9.85 &$ >23.95  $ &$23.72(16)$ &$22.77(13)$ &$22.84(19)$ &26.44 &4.94 &$ >21.71  $ &$ >21.19  $ &$ >19.84  $ \\
      35.44 &4.88 &$ >24.46  $ &$23.75(11)$ &$23.28(14)$ &$23.60(34)$ &35.44 &4.93 &$ >21.44  $ &$ >21.08  $ &$ >19.47  $ \\
      \hline                  
    \end{tabular*}
    \tablefoot{The GRB trigger time is $t_0=55549.686$ MJD. All data
      corrected for $A_{V,\text{host}}=0.12\pm0.01$ mag (SMC).
      \tablefoottext{a}{The duration of the observation.}  }
  \end{minipage}
\end{table}

  \section[Sequences of standard stars]{Sequences of standard stars}\label{app2}

  \begin{minipage}{\textwidth}
    The sequence of reference stars in the field of GRB 091127/SN
    2009nz are taken from \citet{Filgas+11b}. Stars from the 2MASS
    catalogue \citep{Skrutskie+06} are used for the $JHK_\text{s}$
    bands.
  \end{minipage}

  \begin{table}[h]
  \caption{Reference stars in the field of GRB 081007/SN 2008hw.}
  \label{tSTD081007}      
  \centering
  \begin{minipage}{\textwidth}
    \begin{tabular*}{\textwidth}{@{\extracolsep{\fill}}cccccc}
      \hline\hline\noalign{\vspace{0.5\smallskipamount}}
      RA &Dec. &$g\,'$         &$r\,'$         &$i\,'$         &$z\,'$          \\
      {[$^\circ$]}   &[$^\circ$]  &               &               &  & \vspace{0.5\smallskipamount} \\
      \hline\noalign{\vspace{0.5\smallskipamount}}
      339.91132 &$-40.14995$   &$19.334\pm0.013$  &$18.940\pm0.011$  &$18.795\pm0.016$  &$18.658\pm0.017$ \\
      339.91385 &$-40.15378$   &$19.050\pm0.011$  &$18.495\pm0.007$  &$18.274\pm0.012$  &$18.075\pm0.012$ \\
      339.92912 &$-40.17095$   &$14.897\pm0.001$  &$14.399\pm0.001$  &$14.220\pm0.001$  &$14.091\pm0.001$ \\
      339.94805 &$-40.15443$   &$\cdots$          &$19.887\pm0.020$  &$18.788\pm0.016$  &$18.285\pm0.014$ \\
      339.95393 &$-40.11101$   &$19.971\pm0.020$  &$19.147\pm0.011$  &$18.807\pm0.016$  &$18.641\pm0.018$ \\
      339.97304 &$-40.11895$   &$19.890\pm0.019$  &$19.201\pm0.012$  &$\cdots$          &$\cdots$         \\    
      339.97329 &$-40.17987$   &$19.631\pm0.016$  &$18.129\pm0.006$  &$16.949\pm0.005$  &$16.399\pm0.004$ \\
      339.98310 &$-40.12094$   &$19.542\pm0.015$  &$18.780\pm0.009$  &$18.488\pm0.013$  &$18.309\pm0.014$ \\
      339.99130 &$-40.17932$   &$17.947\pm0.005$  &$17.691\pm0.004$  &$17.588\pm0.007$  &$17.517\pm0.007$ \\
      \hline                  
      \end{tabular*}
      \tablefoot{All are observed magnitudes in the AB system.}
    \end{minipage}
  \end{table}

  \begin{table}[h]
  \caption{Reference stars in the field of GRB 101219B/SN 2010ma.}
  \label{tSTD101219B}
  \centering
  \begin{minipage}{\textwidth}
    \begin{tabular*}{\textwidth}{@{\extracolsep{\fill}}cccccc}
      \hline\hline\noalign{\vspace{0.5\smallskipamount}}
      RA &Dec. &$g\,'$         &$r\,'$         &$i\,'$         &$z\,'$          \\
      {[$^\circ$]}   &[$^\circ$]  &               &               &  & \vspace{0.5\smallskipamount} \\
      \hline\noalign{\vspace{0.5\smallskipamount}}
      12.22124 &$-34.52946$   &$22.85 \pm0.10 $  &$19.87 \pm0.05 $  &$18.365\pm0.019$  &$17.720\pm0.018$ \\
      12.25499 &$-34.54326$   &$19.021\pm0.031$  &$18.006\pm0.011$  &$17.568\pm0.012$  &$17.368\pm0.015$ \\
      12.26985 &$-34.56368$   &$21.02 \pm0.07 $  &$19.519\pm0.034$  &$18.475\pm0.022$  &$18.088\pm0.025$ \\
      12.22871 &$-34.56753$   &$15.350\pm0.003$  &$14.928\pm0.002$  &$14.772\pm0.002$  &$14.674\pm0.003$ \\
      12.24608 &$-34.57123$   &$17.328\pm0.010$  &$17.099\pm0.006$  &$17.017\pm0.009$  &$16.970\pm0.011$ \\
      12.21783 &$-34.57419$   &$20.41 \pm0.06 $  &$19.226\pm0.029$  &$18.663\pm0.026$  &$18.484\pm0.039$ \\    
      12.26934 &$-34.58304$   &$16.972\pm0.007$  &$16.599\pm0.004$  &$16.446\pm0.006$  &$16.398\pm0.007$ \\
      12.25056 &$-34.58787$   &$16.996\pm0.007$  &$15.583\pm0.003$  &$14.623\pm0.002$  &$14.166\pm0.002$ \\
      12.21137 &$-34.59066$   &$18.380\pm0.017$  &$18.173\pm0.013$  &$18.097\pm0.016$  &$18.125\pm0.027$ \\
      \hline
    \end{tabular*}
    \tablefoot{All are observed magnitudes in the AB system.}
  \end{minipage}
\end{table}

  \clearpage

  \section[Blackbody fits]{Blackbody fits}\label{app3}

  \begin{minipage}{\textwidth}
    Here we present the blackbody fits for the analysed GRB-SNe, which
    were utilised to estimate the NIR contribution
    (Sect.\ \ref{s3_NIRcor}).
  \end{minipage}

  \begin{figure}[!h]
  \begin{minipage}{\textwidth}
    \centering
    \includegraphics[width=.89\hsize,clip]{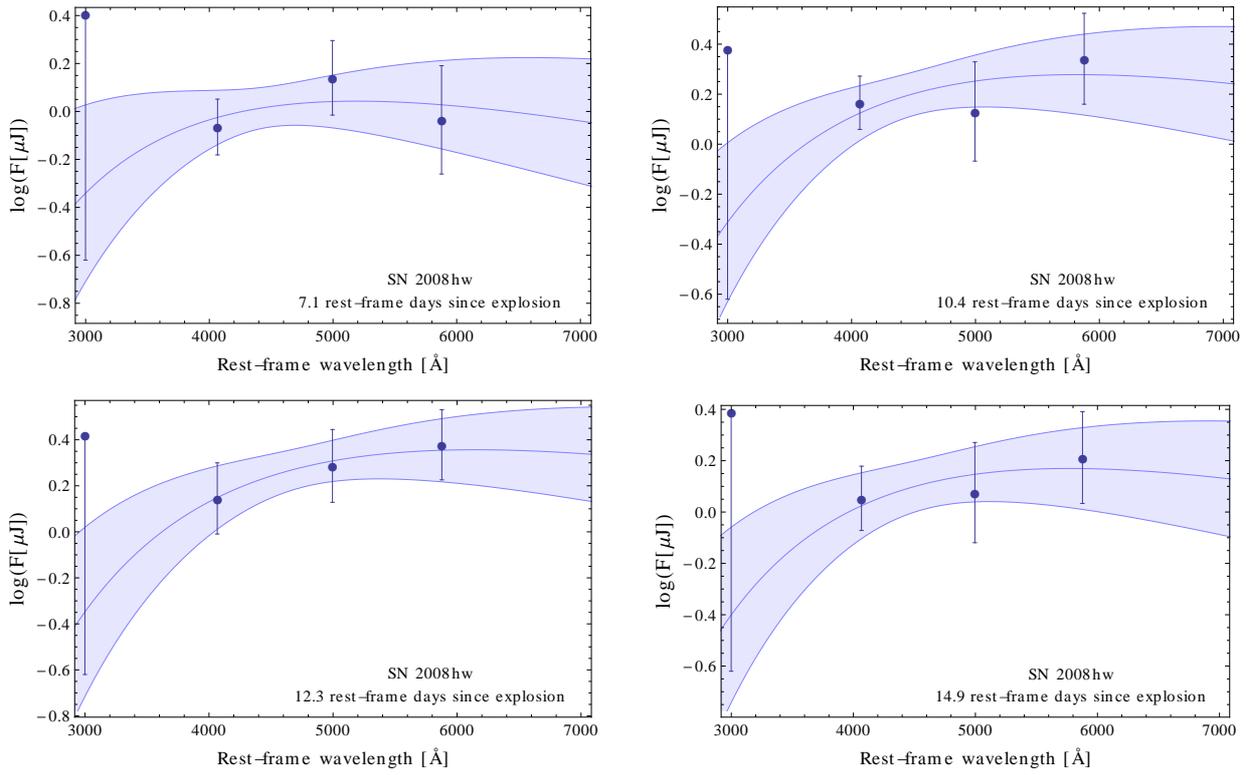}
    \caption{Blackbody fits to the optical photometry of SN
      2008hw. Colour temperatures are about 5,000 K. Points with only
      a lower error bar are upper limits. The blue shaded region shows
      the area between the 1$\sigma$ contours, where the central line
      is the best fit.}
    \label{fBB2008hw}
  \end{minipage}
\end{figure}

\begin{figure}[!h]
  \begin{minipage}{\textwidth}
    \centering
    \includegraphics[width=.89\hsize,clip]{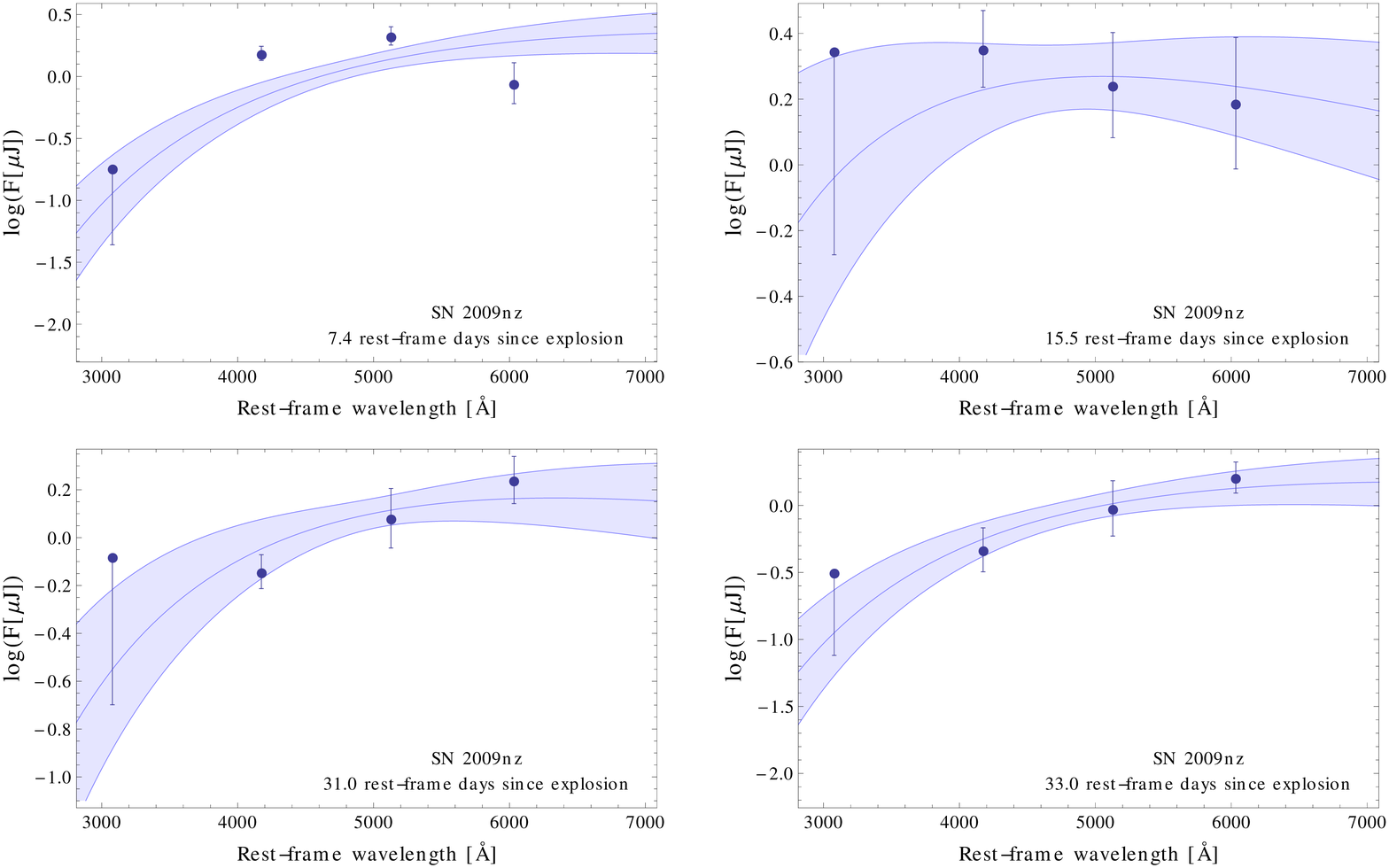}
    \caption{Blackbody fits to the optical photometry of SN
      2009nz. Colour temperatures evolve from $\sim7,000$ to
      $\sim4,000$ K approximately. Point, line, and region coding are
      the same as in Fig.\ \ref{fBB2008hw}.}
  \end{minipage}
  \label{fBB2009nz}
\end{figure}

\begin{figure*}
  \centering
  \includegraphics[width=.89\hsize,clip]{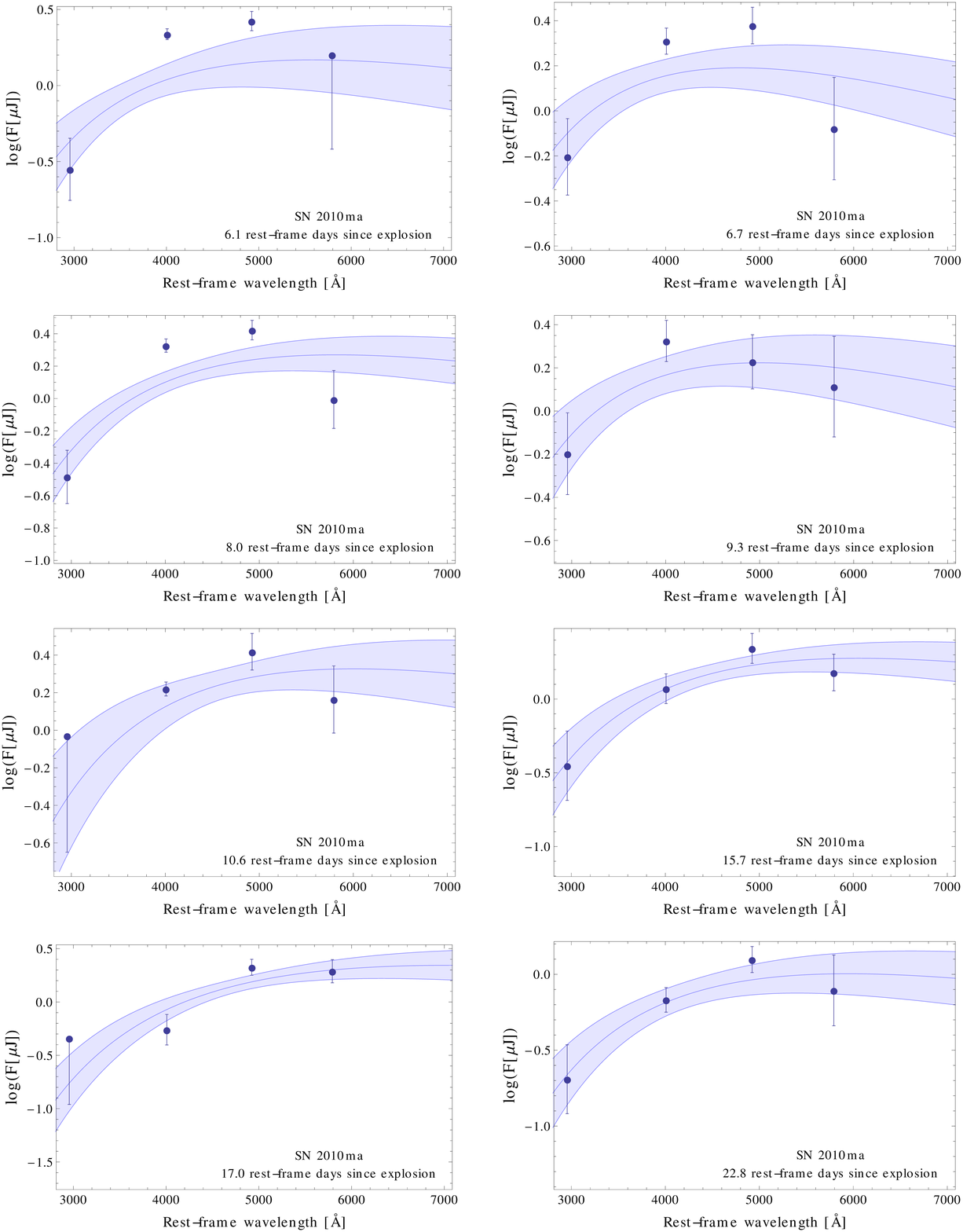}
  \caption{Blackbody fits to the optical photometry of SN
    2010ma. Colour temperatures evolve from $\sim6,000$ to $\sim4,000$
    K. Point, line, and region coding are the same as in
    Fig.\ \ref{fBB2008hw}.}
  \label{fBB2010ma}
\end{figure*}

\end{appendix}

\end{document}